\documentclass[a4paper,11pt]{article}
\usepackage[top=2cm,bottom=2cm,left=2cm,right=2cm,marginparwidth=1.75cm]{geometry}

\usepackage{amsmath,amstext,amssymb,amsfonts,graphics,graphicx,mathtools}
\DeclareMathOperator{\Tr}{Tr}
\usepackage{mleftright} % for '\mleft' and '\mright' macros
\usepackage{times}
\usepackage{caption}
\usepackage{subcaption}
\usepackage{listings}%
\usepackage{multicol,multirow}
\usepackage{tabularx,booktabs}
\usepackage{boites,enumerate}
\usepackage[frozencache,cachedir=.]{minted}
\usepackage{authblk}
\usepackage[flushleft]{threeparttable}

\providecommand{\keywords}[1]{\textbf{\textit{Keywords ---}} #1}
\usepackage{enumitem}
\usepackage{braket}
\usepackage{gensymb}
\usepackage{rotating} % Rotating table
  
\usepackage{hyperref}

\usepackage[numbers]{natbib}
\hypersetup{breaklinks=true}

\newcounter{rowcntr}[table]
\renewcommand{\therowcntr}{\thetable.\arabic{rowcntr}}

\newcolumntype{N}{>{\refstepcounter{rowcntr}\therowcntr}c}

\AtBeginEnvironment{tabular}{\setcounter{rowcntr}{0}}

\title{A digital twin of atomic ensemble quantum memories}

\author[1,*]{Elizabeth Jane Robertson$^\dagger$}

\author[1]{Benjamin Maa{\ss}$^\dagger$}

\author[1]{Konrad Tschernig}

\author[1,2,3,4]{Janik Wolters}

\affil[1]{Institute of Space Research, German Aerospace Center (DLR), Berlin, Germany}

\affil[2]{Institutes of Physics, Technische Universit{\aa}t Berlin, Berlin, Germany}

\affil[3]{Einstein Center Digital Future (ECDF), Berlin, Germany}

\affil[4]{AQLS UG Haftungsbeschränkt, Germany}

\begin{document}

\maketitle

%\fundingInfo{Text}
%\JELinfo{ejlje}
\begin{abstract}
    Accurate performance estimation of experimentally demonstrated quantum memories is key to understand the nuances in their deployment in photonic quantum networks. While several software packages allow for accessible quantum simulation, they often do not account for the loss and noise in physical devices. We present a framework for modeling ensemble-based atomic quantum memories using the quantum channel formalism. We provide a Kraus matrix representation of several experimentally implemented state-of-the art quantum memories and give an overview of their most important performance metrics. To showcase the applicability of this approach, we implement a memory-assisted quantum token protocol within our simulation framework. Our digital twin model is readily extensible to other memory implementations and easily compatible with existing frameworks for performance simulation of experimental quantum networks.
\end{abstract}


\renewcommand\thefootnote{}
\footnotetext{$^\dagger$ The authors contributed equally to this work.}
\footnotetext{* Corresponding author, Email: elizabeth.robertson@dlr.de}

\section{Introduction}\label{sec:intro}
The vision of a quantum internet \cite{kimble_quantum_2008} has driven quantum technology research over the last decades. The backbone of such a network is entanglement distribution between remote network nodes that are connected via photonic quantum channels \cite{duan_long-distance_2001}. However, optical losses in fibers severely limit the possible scale of entanglement distribution and create the need for quantum repeaters. In this context, quantum memories are critical components of any repeater-based transfer of quantum information. Their application in buffering, synchronizing and conditioning of photonic quantum information requires an in-depth knowledge of the inner workings of quantum memories \cite{lvovsky_optical_2009,shinbrough_broadband_2023}. Equally as important is the ability to understand the performance of different quantum memory platforms, experimental implementations and compatibilities with other photonic resources, all within the context of multi-layered network architectures.

Theoretical investigations of ensemble-based quantum memories on the one hand are well-established \cite{gorshkov_universal_2007,novikova_optimal_2007} but often remain very specialized in their scope. On the other hand, recent years have shown remarkable progress in the development of tool-kits for photonic quantum simulation. Prominently, \textit{Strawberry fields} \cite{killoran_strawberry_2019}, \textit{Perceval} \cite{heurtel_perceval_2023} and \textit{Piquasso} \cite{kolarovszki_piquasso_2025}  have emerged as tools for Gaussian- and discrete Boson sampling and photonic quantum computing. In addition, tools for simulating open quantum systems \textit{QuTiP}\cite{lambert_qutip_2024} and gate-based computing \textit{Qiskit}\cite{javadi-abhari_quantum_2024} with extension module for optics \textit{SOQCS}\cite{osca_soqcs_2024} have been published. However, there is a lack of open source, close-to-experiment plug-and-play software packages to simulate real quantum communication devices taking into account noise, loss and decoherence processes which requires significant manual customization in presently available simulation frameworks. 

In this work we develop a general model of atomic ensemble-based quantum memories, taking loss-, noise- and decoherence processes into account, and demonstrate the plug-and-play software implementation in several test-scenarios. First, we describe the basic working principle of ensemble-based quantum memories and introduce their most important performance characteristics \ref{sec:quantum_memories}. Second, after introducing the description of a quantum memory within the channel formalism \ref{sec:Channel_Formalism} we demonstrate the functionality of the simulation and apply it in a Mach-Zehnder configuration. Furthermore, we introduce simulation building blocks as digital twins of state-of-the-art vapor memories which allows for straightforward performance comparisons. Finally, we apply the digitally twined memories to a quantum token protocol and benchmark the memory performances against the security of the protocol \ref{sec:examples}.

\section{Quantum Memories}\label{sec:quantum_memories}
Fundamentally, a memory is a device that stores some type of information encoded in a propagating or short-lived state. The information is transferred to a storage medium by encoding it in a long-lived storage state and retrieved at a later time with a certain efficiency and fidelity. The term efficiency refers to the probability of a successful memory operation versus the number of trials. The fidelity of a memory is defined as the similarity between the input and the retrieved information. In the case of quantum states to be stored the re-encoding of information in a storage medium is fundamentally difficult as it entails a projective measurement of the quantum state which changes the properties of the state itself. Specifically, an unknown quantum state cannot be reconstructed from a single measurement. Crucially, this means that storing quantum information requires storing the \textit{entire} state and not only its projections onto some measurement basis during the storage process. The challenge in constructing such a system lies in the fundamental impossibility to create identical copies of arbitrary quantum states\cite{wootters_single_1982}. 

\begin{figure*}[h]
    \centering
    \includegraphics[width=1\textwidth]{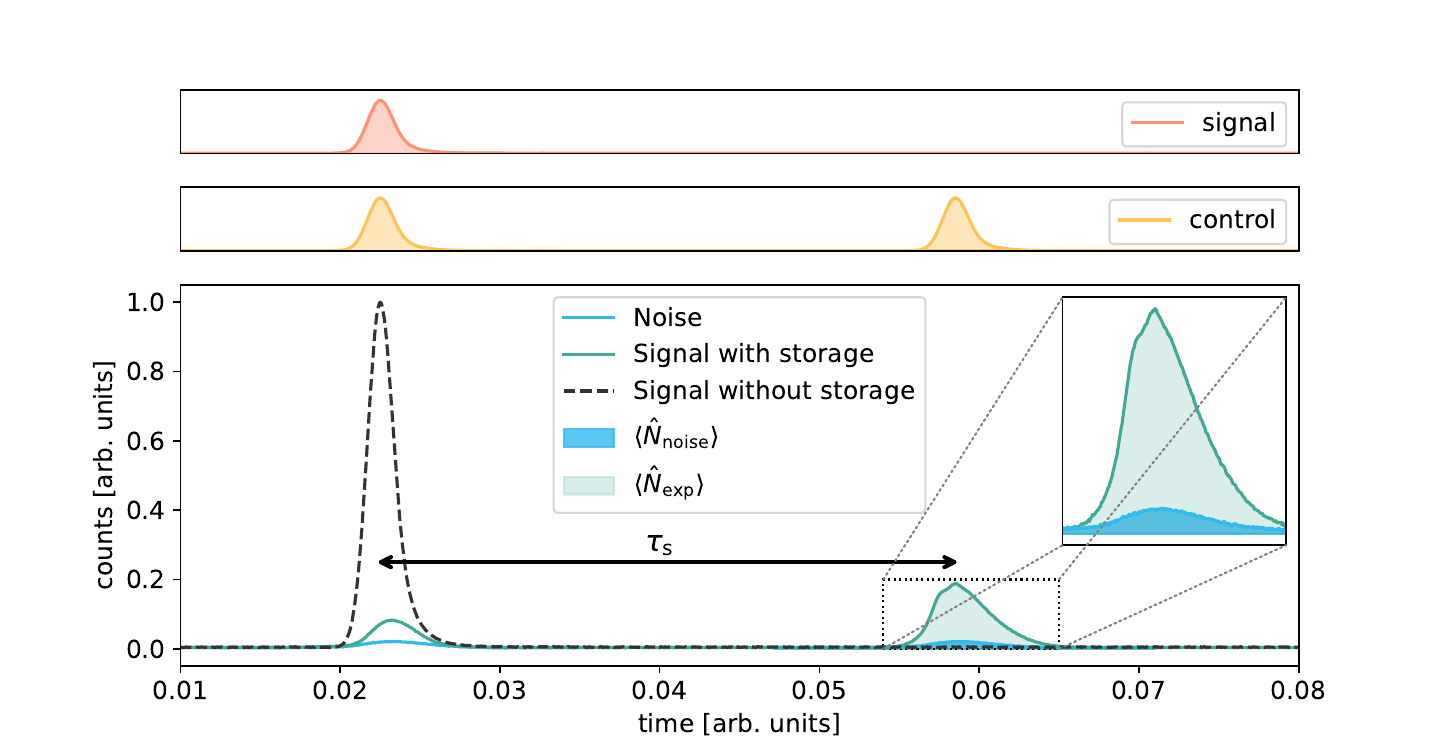}
    \caption{A typical photon arrival time histogram for an optically controlled memory experiment. The upper panels show the envelopes of the signal and the control field fed into the memory. The overlap of the control pulse with the signal pulse within the medium at time 0.22 maps the signal pulse to a collective spin-(orbital) wave. The experimental result (bottom panel) shows some leakage at this point in time from unstored light. After the storage time $\tau_\mathrm{s}$ the control pulse is re-applied and the signal field is retrieved.}
    \label{fig:storage_trace}
\end{figure*}

In this work we focus on photonic quantum memories in particular, i.e. memory devices that store information encoded in states of light. The most straight-forward way of storing light is in a fiber loop or Herriot cell. An incoming state of light is guided onto a long or, in the case of a fiber loop, closed light path that delays the transmission of the state. These implementations usually provide high fidelity and broadband operation but come with fixed read-in and read-out intervals \cite{fook_lee_fiber_2024}. This work concentrates on \textit{optically controlled} ensemble-based memories. In these implementations a light field is used to exert control over the interaction between the light field to be stored and some atomic ensemble that serves as storage medium. During the read-in process the photonic state is mapped onto a coherent light-matter state, sometimes referred to as spinwave or (dark state) polariton under an ideally unitary (time-reversible) transformation. This allows one to reverse the read-in process after the storage time $\tau_\mathrm{s}$ and retrieve the initial photonic state. During the storage time the state undergoes some decoherence processes that heavily depend on the specific implementation. However, since both the read-in and read-out processes are (ideally) unitary the quantum coherence between the initial state and the retrieved state is conserved. An example of the storage and retrieval of a coherent state from an ensemble vapor memory can be seen in Figure \ref{fig:storage_trace}. The signal pulse (red) is mapped onto the spinwave by applying the first control pulse (orange). After time $\tau_{\mathrm{s}}$, a second control pulse is sent into the atomic ensemble, and the state is retrieved from the memory (light blue). 

There is a wide variety of different memory implementations with different properties that determine their suitability for application.
In the next section, we describe the most important key characteristics on the basis of which ensemble quantum memories are compared.

\subsection{Characteristics and figures of merit}\label{sec:characteristics}

\subsubsection{Efficiency}
\label{subsec:efficiency}
\begin{figure*}[h]
    \centering
    \subfloat{\includegraphics[width=0.4\textwidth]{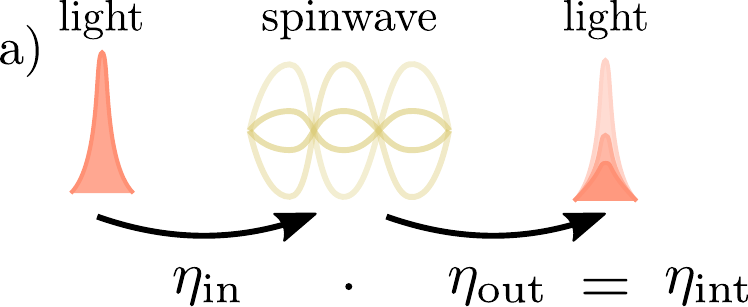}}
    \qquad
    \subfloat{\includegraphics[width=0.4\textwidth]{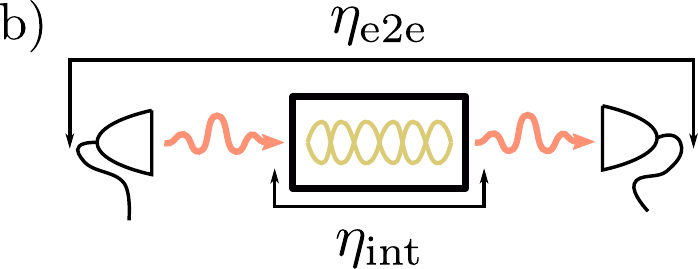}}
    \caption{Internal and external efficiencies of a quantum memory. a) The incident light field gets stored in the spinwave with efficiency $\eta_\mathrm{in}$ and retrieved with efficiency $\eta_\mathrm{out}$ with a combined internal efficiency of $\eta_\mathrm{int}$. b) The end-to-end efficiency $\eta_\mathrm{e2e}$ includes the optical transmission through the setup.}
    \label{fig:efficiencies}
\end{figure*}

The efficiency of a quantum memory describes the probability of retrieving a photon from the memory after a certain time, i.e. generally the memory efficiency is a function of the time between read-in and read-out. 
In some cases it is useful to describe the efficiency of the read-in and the read-out processes separately. 
The read-in efficiency $\eta_\mathrm{in}$ describes the probability of mapping a photon onto a coherent matter state when applying an optical control field. In the literature, it is sometimes referred to as storage efficiency.  Respectively, the read-out efficiency $\eta_\mathrm{out}$ gives the probability of retrieving the photon field from the matter state at a later time. Combining the two, we define the internal efficiency $\eta_\mathrm{int}=\eta_\mathrm{in}\cdot\eta_\mathrm{out}$ as the probability of successful read-in and read-out, not accounting for optical losses before or after the storage medium. Taking these into account, we define the end-to-end efficiency $\eta_\mathrm{e2e}$ which includes technical losses from the setup, $\eta_{\mathrm{e2e}} = \eta_{\mathrm{trans}} \cdot \eta_{\mathrm{int}}$. A depiction of the different efficiency definitions can be found in Figure \ref{fig:efficiencies}. Typical experiments achieve end-to-end efficiencies on the order of 1\%-35\%.

\subsubsection{Storage time}
The time between read-in and read-out of a photonic state is generally referred to as storage time $\tau_\mathrm{s}$. During storage, the medium is subject to experiment specific decoherence processes which limit the storage time of the memory.  In most cases homogeneous effects are dominant (i.e. all constituents of the storage medium undergo the same decoherence) and the internal efficiency exponentially decays with the storage time. Consequently, memory experiments report the 1/e storage time, also known as the memory lifetime, as a metric which can be used to compare different memory implementations. Typical storage times, depending on the memory protocol, range from nanoseconds to hundreds of microseconds.

Different ensemble based quantum memories use different memory protocols, that is, there is a different underlying physical process which controls the storage and retrieval process. A useful way to classify memories is to distinguish between memory implementations with fixed read-out intervals and \textit{on-demand} read-out. Protocols that rely on the periodic rephasing of ensembles (e.g., atomic frequency combs - AFC) generally don't operate on demand; however, they may be modeled using the approach below. It is left to the user to take care that read-out only occurs at the rephasing period.
Another important metric that, together with the storage time, defines the memory's throughput is the memory downtime i.e. the time that is required to reinitialize the memory after a storage operation. The memory downtime is mostly limited by state preparation of the storage medium e.g. optical pumping.

\subsubsection{Operational wavelength and bandwidth}
The operational wavelength of the memory depends on the atomic medium selected and its optical transitions. Most prominently, alkali metal vapors are used as storage media due to their strong and spectrally well-separated dipole transitions in the near infrared. The frequency of the input photon state has to be matched to the memory operation wavelength which is, within some detuning, determined by the chosen optical hyperfine transition. Detuning of the input signal from the optical transition can be beneficial in some memory implementations but heavily depends on the specific case.

\begin{figure*}[h]
    \centering
    \subfloat{\includegraphics[width=1\textwidth]{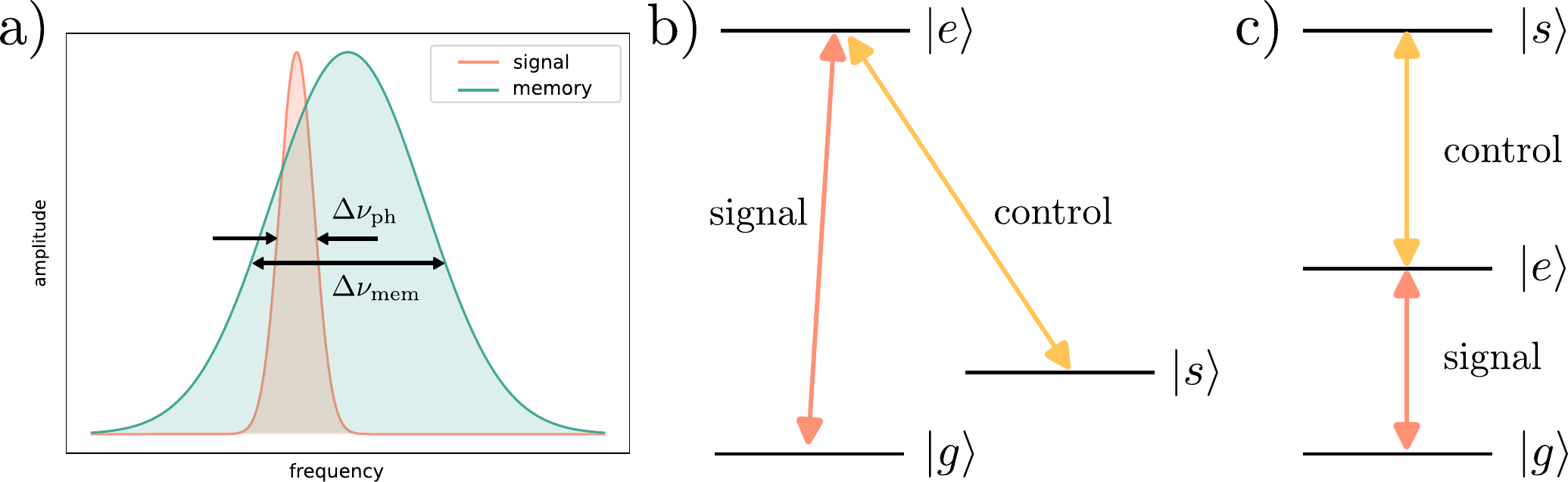}}
    
    \caption{Bandwidth and memory type comparison. a) Spectral overlap of memory bandwidth and signal spectrum. For practical applications both signal wavelength and signal linewidth have to match the spectral properties of the memory. b) A memory in $\Lambda$-type configuration operates between ground- and storage states within one hyperfine manifold. c) Ladder-type memory configurations connect ground and storage state via full orbital transitions.}
    \label{fig:bandwidthandcomparison}
\end{figure*}

The bandwidth of a memory $\Delta\nu_\mathrm{mem}$ determines the spectral width of an input photon $\Delta\nu_\mathrm{ph}$ that can be stored in the memory, see Figure \ref{fig:bandwidthandcomparison}. For Fourier-limited pulses the photon's spectral width is inversely proportional to its pulse width $\tau_\mathrm{pulse}$. Consequently, a short photon pulse to be stored requires a higher bandwidth of the memory. The ratio of memory storage time and shortest possible signal pulse length is called the fractional delay and gives an intuitive picture of the memory's ability to delay a pulse. The time-bandwidth product $B\sim\tau_s/\tau_\mathrm{pulse}$ of a memory is an equivalent measure which is most established in the literature. The product of the memory's time-bandwidth product and its end-to-end efficiency is regarded as the most important figure of merit of quantum memory implementations.

\subsubsection{Noise level}

Memories with unit efficiency are rendered unusable if the number of noise photons in the retrieval mode exceeds the number of actually retrieved photons. Depending on the memory implementation, a plethora of physical mechanisms can influence the noise performance of a memory. Most prominently, optically controlled memories suffer from Four-wave mixing, Raman-noise and fluorescence. In principle, different noise levels can occur during read-in and read-out, which we account for in the simulation. The signal-to-noise ratio (SNR) is the most-used metric in the literature. It refers to the number of photons that are retrieved from the memory compared to the average number of detected noise photons during the retrieval: 
\begin{equation}
\text{SNR}=(\langle \hat{N}_\mathrm{exp}\rangle-\langle \hat{N}_\mathrm{noise}\rangle)/\langle \hat{N}_\mathrm{noise}\rangle 
\end{equation}
where $\langle \hat{N}_\mathrm{exp} \rangle$ is the average number of total counts that are detected within the retrieval window, including the average number of noise photons $\langle \hat{N}_\mathrm{noise} \rangle$. These parameters can be extracted from experimental data, such as the arrival time histogram shown in Figure \ref{fig:storage_trace}. Another metric often used in the literature is the unconditional noise figure $\mu_1$. It relates the SNR to the number of input photons $\langle N\rangle_\mathrm{in}$ via:

\begin{equation}
\mu_1=\frac{\langle N\rangle_\mathrm{noise}}{\eta_\mathrm{int}}=\frac{\langle N\rangle_\mathrm{in}}{\mathrm{SNR}}.
    \label{eq:mu1}
\end{equation}
While the SNR is a figure that depends on the number of input photons the noise figure characterizes the number of noise photons alone. Importantly, we can use \ref{eq:mu1} to calculate the average number of noise photons from the $\mu_1$ parameter which we will use in the simulation section (see Section \ref{sec:simulation}). The $\mu_1$ values in experiments are on the order of $10^{-2}$ to $10^{-6}$.

\subsubsection{Single-photon operation}

Interfacing existing memory implementations with true single-photon sources poses a significant challenge. The biggest problem lies in the bandwidth mismatch between single photons and memories. Quantum memories based on atomic ensembles typically exhibit bandwidths in the few MHz- to sub GHz-range. On the other hand, the most researched single-photon sources are based on solid-states (quantum dots, SPDC in bulk crystals etc.) and offer photon bandwidths from hundreds of MHz to hundreds of GHz. In addition to the bandwidth matching, the operational wavelengths of the single photon source and the memory have to be aligned. Considering the limited number of available dipole transitions in atomic ensembles, this is a highly non-trivial task. Moreover, for every type of photon source the read-in process has to be matched to the specific photon temporal envelope. Probabilistic photon sources come with the additional complexity of synchronizing the memory operation to the herald of a photon creation. Together with the numerous sources of background emission in the generation of single photons this places very stringent requirements on experimental implementations of true single photon storage. Previously, single photons from an SPDC sources have been coupled to a $\lambda$- type memory \cite{buser_single-photon_2022, tsai_quantum_2020, michelberger_interfacing_2015}. A ladder-type memory has been used to store and synchronize single photons from a room-temperature Four-Wave mixing source \cite{davidson_single-photon_2023}. More recently, single photons from semiconductor quantum dots have been stored and retrieved from room-temperature ladder-type atomic vapor memories \cite{thomas_deterministic_2024,maas_-demand_2025}.

\subsubsection{Fidelity}

The fidelity of a quantum memory describes its ability to retrieve an unaltered quantum state from the memory. In its most general form it can be written as a comparison between two quantum states described by the density matrices $\hat{\rho}$ and $\hat{\sigma}$: \begin{equation}
    \mathcal{F}=\left(\mathrm{tr}\sqrt{\sqrt{\hat{\rho}}\hat{\sigma}\sqrt{\hat{\rho}}}
\right)^2
\end{equation} . For pure states, the fidelity reduces to the overlap between the two states $\mathcal{F}=\left|\langle{\psi_e}|\psi_l\rangle \right|^2$.  Even though the fidelity is arguably the most important memory performance metric, it is, paradoxically, rarely reported in the literature. The reason is that the measurement of $\mathcal{F}$ requires two full quantum tomography measurements of the in- and output states $\rho$ and $\sigma$, which is notoriously difficult. In this regard, our digital twin model enables the inference of the fidelity performance from memory parameters that are much easier to access experimentally. The main effects that lowers the fidelity of a memory is the loss of photons and the generation of noise photons during read-out. In addition to that, the memory operation may imprint phase noise onto the state or distort the single photon wavepacket which generally results in an overall lower fidelity of a memory. The literature provides some results of memory experiments that measure the interferometric visibility of stored and un-stored light \cite{reim_towards_2010,saglamyurek_coherent_2018} which can serve as an upper bound on the achievable memory fidelity. We include a simulation of state fidelities after a memory operation in the Appendix \ref{app:Fidelity}.

\subsubsection{Operational simplicity and multi-mode capacity}
Considering that a new infrastructure is required for the mass deployment of quantum memories, the operational simplicity of the technology is critical for long-range entanglement distribution. Compatibility with existing telecom fiber networks is desirable. Ideally, quantum memories can be micro-integrated or miniaturized \cite{mottola_optical_2023, jutisz_stand-alone_2025,liu_millisecond_2025} and operated without cryogenic cooling. Low energy consumption and robustness against vibrations, temperature changes, radiation etc. is needed to allow for memory operation in space. Beyond these requirements, the ability to store more than one light mode in the same device (at the same time) is referred to as multi-mode capacity.\cite{ li_multicell_2021,messner_multiplexed_2023,zhang_realization_2024}. It will be of critical importance in the scaling of quantum network infrastructures.

The atomic transitions that can be addressed during the storage process depend on selection rules and the light polarization. Therefore, ensemble-based memories are not polarization agnostic in general and each orthogonal polarization requires a separate memory mode. An important consequence of this fact is the requirement of at least two memories or memory modes to store a full polarization qubit \cite{namazi_ultralow-noise_2017}. Some of the memory implementations from the literature accept linear polarization and some only work with circular polarized light, which we have noted in the overview Table \ref{tab:memory_triggertimes}.

\subsection{Memory configurations and protocols}

Quantum memories can operate in diverse regimes depending, for example, on their signal detuning from optical transitions, optical depth of the atomic ensemble or Rabi frequency. The different operating regimes result in different characteristics as discussed in the previous section.  For detailed explanations of the different working regimes, we refer the reader to a review article on the topic \cite{shinbrough_broadband_2023}. In this section, we give a very brief explanation of the most fundamental differences between $\Lambda$-type and ladder-type configurations (see Figure \ref{fig:bandwidthandcomparison}), which aims to help to understand the performance differences between memories that can be seen in Table \ref{tab:memoryperformances}.

Fundamentally, the process of storage and retrieval relates a (collective) atomic ground state to a storage state via an intermediate state. In $\Lambda$-type configurations the storage state is within the same orbital level as the ground state only separated by the hyperfine splitting. The storage process results in a spin coherence between two states that are energetically separated by a few GHz (in the case of alkali ensembles). The wavevector difference $\Delta k=|k_\mathrm{s}-k_\mathrm{c}|$ between the signal and the control field that connect the ground state and the storage state determines the coherence length of the spinwave. Consequently, $\Lambda$-type memories with small wavevector differences show the longest reported storage time (e.g. \cite{katz_light_2018,wang_field-deployable_2022}). However, the small energy separation between signal and control fields comes with increased technical complexity due to spectral filtering and introduces noise channels to the system (e.g. Four-wave mixing, Raman noise). It can be seen from the overview table \ref{tab:memoryperformances} that $\Lambda$-type memories have the lowest SNR values. On the other hand, ladder-type memories (sometimes referred to as $\Xi$-type) bring the advantage of almost noise-free operation due to the large energy separation between ground state and storage state. The drawback of this configuration lies in the short storage times (order of tens of nanoseconds) compared to $\Lambda$ memories.

 Within the broad categories of ladder and lambda memories, different operating parameters (signal detuning, control power ect.) result in different underlying physical processes mediating the storage process. These different operating regimes are called memory protocols, and determine the specific behavior of the memories. Off-resonant cascaded absorption (ORCA) and fast ladder memory (FLAME) are typical memory protocols for ladder memories, while Raman-electromagnetically induced transparency (EIT) and Raman-Autler-Townes splitting (ATS) usually refer to operating regimes in Lambda systems. We list the memory scheme and the protocol used in Table \ref{tab:memory_triggertimes}. For more details on the specific operation conditions and the memory protocols we refer to the review on broadband vapor memories \cite{shinbrough_broadband_2023}.

Some characteristics such as operational simplicity, multi-mode capacity and single photon operation, are measures derived from demonstrated experimental implementation and are descriptive. Consequently they are not explicitly accounted for in the mathematical models we derive in the following section.  
The figures of merit, efficiency, storage time, operational wavelength and bandwidth,  noise level, fidelity and polarization represent measurable parameters on the basis of which we can compare different memory implementations. In the next section we introduce the channel formalism and explain how the memories' figures of merit are implemented in the simulation.

\section{Channel Formalism}\label{sec:Channel_Formalism}
To model different implementations of ensemble-based quantum memories, we define a general model for an ensemble vapor memory.  First, we present the Hilbert space of the used formalism, before considering the memory process, and the experimental noise and losses. In Section \ref{sec:simulation} we explain the implementation details of our code and demonstrate some applications. 

\begin{figure}[h]
    \centering
    \includegraphics[width=\linewidth]{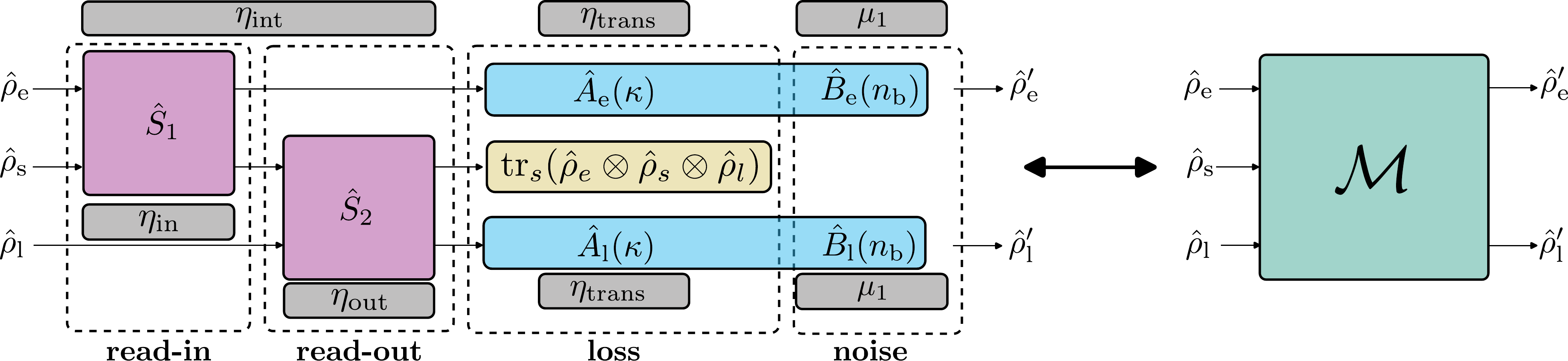}
    \caption{Depiction of the quantum memory operation in the quantum channel formalism. $\hat{\rho}_e, \hat{\rho}_l $ are the input states to the memory, corresponding to the states in the early and late time bin. $\hat{\rho \prime}_e, \hat{\rho \prime}_l$ are the early and late states after the memory operation. $\hat{\rho}_s$ is the state representing the spinwave, which is traced out after the memory operation. Every colored box corresponds to a set of Kraus operators with annotated parameters from memory experiments. The parameters in brackets correspond to the variables in the code. The memory box on the right serves as an abbreviated version for better readability.}
    \label{fig:flowchart}
\end{figure}

\subsection{Hilbert space}
 We choose the Fock basis for the model and denote the path, polarization and time bin of the photons as mode indices. As such, the state $\ket{n}_{\{\mathrm{X,p,t}\}}$ corresponds to $n$ photons in path X, with polarization $p$, in the time bin $t$. Mode parameters which are not relevant to the system being modeled are then omitted i.e. if all photons in the system have the same polarization, we write the state as $\ket{1}_{\mathrm{a,e}}$.
 
As a central tenet of memories is the delay of information from an earlier to a later time, the time-bin encoding of quantum information is common in the memory community. The Hilbert space acted on by the model is the joint space between two Hilbert  spaces $\mathcal{H}_{e} \otimes \mathcal{H}_l$, where the input states of the memory are labeled with index $e$ and the retrieved states in the late time bin with index $l$, respectively. 

% Multimode memories (see e.g. [REFERENCES] can be modeled as multiple memories in parallel.
Memories are polarization-specific, and are often (although not exclusively) single path mode devices. Thus, we can simplify the notation down to the time-bin indices, i.e.  $\ket{1}_e$. The joint state of say a single photon in the early time bin $\ket{1}_{e}$ and no photon in the late time bin $\ket{0}_l $ is then written $\ket{1}_e \otimes \ket{0}_l = \ket{1_e, 0_l} = \ket{10}$

In this work we present the operation of the memory using the channel formalism, that is states shall be written in density matrix form:
$\hat{\rho}_e = \sum_n p_n\left (\ket{\psi_n}\bra{\psi_n}\right )_e$, and the operators representing the memory are written in the Kraus form: 
$\sum_i K_i^{\dagger}K_i = 1$, where 1 is the identity.

A perfectly operating generalized memory stores an incoming photonic state $|n\rangle_e$ and perfectly returns that state at a later time.

%$$\hat{S}\left(\left (\ket{}\bra{\Psi}\right )_e\otimes\left (\ket{0}\bra{0}\right )_l\right)= \left ( \ket{0}\bra{0}\right )_e\otimes \left ( \ket{\Psi}\bra{\Psi} \right )_l$$ 

$$\hat{S}\hat{\rho}_e \hat{S}^{\dagger}= \hat{\rho}_l $$
where $\rho_e = \rho_l$ and $\hat{S}$ one whole memory operation (read-in and read-out). However, physical memories are not perfect systems; the storage- and retrieval processes have limited efficiency, a single photon incident in the late time-bin may be stored, losses are present in the optical setup, and noise photons are present. Figure \ref{fig:flowchart} depicts how each of these effects is represented in our model. In the following sections, we deal with each of these effects in turn in greater detail.
% We shall represent these two processes with two operators $\hat{S}_{\mathrm{in}}, \hat{S}_{\mathrm{out}}.$

\subsection{Imperfect read-in and read-out}
As mentioned in section \ref{subsec:efficiency} the memory process consists of two operations, the mapping of an incoming state to a spin wave (read-in), and the retrieval of the state from the atomic ensemble (read-out).

The read-in operator of an ensemble-based quantum memory, $\hat{S}_{\mathrm{in }}$, transfers the incoming state $\hat\rho_e$ to a state $\hat\rho_s$, representing the internal state of the memory (spinwave). As the storage process is often imperfect, we model the storage operator as a beamsplitter with transmittivity, $t_{\mathrm{in}} = 1 - \eta_{\mathrm{in}}$, where $\eta_{\mathrm{in}}$, is the read-in efficiency, defined as the fraction of photons successfully transferred to the spinwave. The leaked photons correspond to the photons remaining in the early time bin after the storage operation $\hat{S}_{in}$. 

The read-out operator, $\hat{S}_{\mathrm{out}}$, is a beamsplitter acting on the space $\mathcal{H}_s\otimes\mathcal{H}_l$, and controls the number of photons which are transferred from the spinwave to the late time-bin. The transmissivity of the $\hat{S}_{\mathrm{out}}$ is  $ t_{\mathrm{out}} = 1- \eta_{\mathrm{out}}$,  where $\eta_{\mathrm{out}}$, is the read-out efficiency. Since the storage state $\rho_{\mathrm{s}}$, is not actively monitored, it is treated as the environment and traced out after the read-in and read-out operations.   

Formally, the read-in operation corresponds to the unitary operator 
\begin{equation}  
S_{\mathrm{in}} = e^{\zeta (e s^\dagger-e^\dagger s )},
\end{equation}

where $\kappa$ is the coupling strength between the early time-bin mode and the memory storage mode. In practice this leads to a linear transformation of the mode operators
\begin{equation}
    S_{in}: (e^\dagger \\ s^\dagger) \rightarrow  \begin{pmatrix}
\sqrt{t_{\mathrm{in}}} & \sqrt {r} \\
-  \sqrt r & \sqrt{t_{\mathrm{in}}}
\end{pmatrix} * (e^\dagger \\ s^\dagger),
\end{equation}

where $r$ is the beamsplitter reflectivity $r = 1  - t_{\mathrm{in}}$ and $e^\dagger$ and $s^\dagger$ are the creation operators of the early and storage spaces. The coupling $\zeta$ relates to the beamsplitter parameters via $\cos(\zeta) = t$ and $\sin(\zeta) =r$.
 This form of the unitary acts on the creation and annihilation operators of the early and late states, however as the input states are in the full matrix representation of the Fock basis the beamsplitter must also be written in the same form:

\begin{equation}
    \widehat{S}_{\mathrm{in}} = \sum^{\infty}_{N= 0} \sum^{N}_{n,m= 0} S^{(N)}_{m,n} \ket{N-m}_e\ket{m}_s\bra{N-n}_e\bra{n}_s,
\end{equation}
where the matrix element $S^{(N)}_{m,n}$ (dropping the $\mathrm{in}$ index for readability),  is given by: 

\begin{equation}
    S^{(N)}_{m,n} = \frac{\sqrt{(N-n)n!}}{\sqrt{(N-m)m!}}\sqrt{t_{\mathrm{in}}}^{N-2n} \left( \frac{\sqrt r}{\sqrt t_{\mathrm{in}}} \right )^{m-n}P^{(m-n, N-n-m)}_{n}(t_{\mathrm{in}}-r),
\end{equation}

 where $P^{(a,b)}_{n}(x)$ are the Jacobi polynomials. A derivation of this result can be found in the supplemental document.
 As it is often experimentally infeasible to determine the read-in and read-out efficiencies independently of one another, it is common practice not to report $\eta_{\mathrm{in}}$ and $\eta_{\mathrm{out}}$ individually, but rather the total internal efficiency $\eta_{\mathrm{int}} =\eta_{\mathrm{in}}\eta_{\mathrm{out}} $. For the memories presented in this paper, we assume the read-in and read-out efficiency to be the same, $\eta_{\mathrm{in}} = \eta_{\mathrm{out}} = \sqrt{\eta_{\mathrm{int}}}$, but the code easily allows for the treatment of unbalanced read-in and read-out efficiencies.

\subsection{Loss and Noise} \label{sec: Loss and Noise}
To represent further losses in the channels and noise we treat the early and late states after the beamsplitters as inputs into a lossy thermal noise channel. To find a Kraus operator description, one decomposes the joint thermal noise channel into a pure loss channel of transmitivity $\tau = \kappa/G $  followed by a quantum limited amplifier with gain $G = 1 +(1-\kappa)\bar{n}_{\mathrm{B}}$, where $\kappa$ is the beamsplitter transmissivity and $\bar{n}_{\mathrm{B}}$ is the mean thermal photon number of the noise channel \cite{gagatsos_bounding_2017}. This results in the following Kraus matrices for the pure loss channel $A_l$ and the amplifier $B_k$.

\begin{equation}
\hat{A}_l =p_l
\tau^{\frac{\hat{n}}{2}} \hat{a}^l,
\end{equation}

\begin{equation}
\hat{B}_k = q_k\hat{a}^{\dagger k} G^{-\frac{\hat{n}}{2}}.
\end{equation}

where $p_l = \sqrt{\frac{(1-\tau)^l}{l!}}, q_k = \sqrt{\frac{1}{k!} \frac{1}{G}\left(\frac{G-1}{G}\right)^k} $, $\hat{n}$ is the number operator and $ a^{({\dagger})}$ are the creation (annihilation) operators. Thus, an input state $\rho_0$ then transforms according to $\rho =  \sum_{k,l =0 }^{\infty} \hat{B}_k\hat{A}_l\rho_0 \hat{A}_l^{\dagger}\hat{B}_k^{\dagger}$, where $k,l$ are indicies to sum over the full fock space. In a truncated fock space, the sum does not go to infinity but to truncation $n$.  In this form, the first the beam splitter operation representing loss $\hat{A}$, is applied before the gain operator. This is convenient as it lets us control the loss of $\rho_0$ with the transmissivity $\kappa$. This is analogous to losses through the experimental setup, after the storage operation and therefore the choice of $\kappa = \eta_{\mathrm{trans}}$ is appropriate. By setting $\kappa = \eta_{\mathrm{trans}}$ and applying the definition of $\mu_1$ \ref{eq:mu1} we can give an analytical formula for $\bar{n}_{\mathrm{B}}$ : 

\begin{equation}
   \bar n_B= \frac{\langle n\rangle_\mathrm{noise}}{1-\eta_\mathrm{trans}}=\frac{\mu_1\eta_\mathrm{int}}{1-\eta_\mathrm{trans}}.
\end{equation}

%In real experiments, loss and gain do not simply occur after one another but rather simultaneously, i.e., noise photons generated may then be lost to the environment - a case which is not represented in the channel description above, as first the beam-splitter is applied, then the noise. The consequence is that there is no immediate analytical form relating $\mathrm{SNR}$ to the average photon number of the thermal lossy channel ($\bar{n}_{\mathrm{B}}$). We fit $\bar{n}_{\mathrm{B}}$ so that our model provides photon statistics that match those measured in memory experiments. The fit values for $n_{\mathrm{B}}$ can be found for \ref{tab:memoryperformances}, and the fit method is presented in the Appendix. \ref{app: fitting}.  We assume the noise is the same for both read-in and read-out operations.

In treating the noise photons as a quantum limited amplifier, the computational complexity of the simulation increases. The truncation behavior of the simulation is presented in Appendix \ref{app:truncation}. We set the truncation $n = 5$ for simulations with coherent states and the truncation $n = 3$ for single photons unless otherwise stated. To simulate systems with higher input photon numbers, i.e. coherent states for $\ket{\alpha}$ where $ |\alpha|^2 > 1$, the simulation truncation should be increased so that the output state converges.

\section{Simulation}\label{sec:simulation}
The model presented above represents a number of simplifications of the underlying physics, which may affect the accuracy of memory performance simulations. In Section \ref{sec:initalization} we discuss both memory and input state initialization,  the checks carried out to ensure compatibility between the two, as well as the limitations of these checks. In this section, further details of the operation of the memory channel query and an example query is shown.  
The use of the memory simulation in a Mach-Zehnder Interferometer experiment and in a quantum token protocol is presented in Section \ref{sec:examples}. An example implementation, can be found at \cite{robertson_notitle_nodate}. To initiate the memory, the user must define several parameters of the memory itself and ensure the properties of the state which is to be stored match the properties of the memory. We first discuss the initialization of the memory, before considering the initialization of the input state. 

\subsection{Memory Initialization}\label{sec:initalization}
To simulate different atomic vapor memories, there are several properties which must be defined by the user upon initialization of the memory; these are, the memory type, the storage time, and the truncation of the internal memory state, and are further discussed below.  An example initialization of the memory is given in code listing \ref{lst:Initalization} where a memory is initialized  with a storage time of $\tau_s = 1\ \mu s$, and with a truncation bound of $n = 5.$  

%[caption=The initalization of 'Lambda895' memory as demonstrated. The internal state of the memory is then initialized to be the vaccum state.,label=lst:Initalization, basicstyle=\fontsize{8}{10}\selectfont\ttfamily]

\begin{listing}
\begin{minted}[framesep=2mm]{python}
mem.set_param('memory_type', 'Lambda895')
mem.set_param('storage_time', 1e-6)
mem.set_param('memory_truncation', 5)
mem.send_params()
mem_state = mem.state_init()[0]
\end{minted}
    
\caption{The initalization of 'Lambda895' memory as demonstrated. The internal state of the memory is then initialized to be the vacuum state.}\label{lst:Initalization}
\end{listing}

The first initialization parameter is the \lstinline{memory type}, defining which memory should be used in the simulation. This is set using the \lstinline{memory_type} parameter which takes a string as an argument. To access different memories, the user should provide the name of a memory class, which represents the different memory experiments. The names of the different classes (and thus memories) available can be found in Table \ref{tab:memoryperformances}. By setting the memory type using the command \lstinline{mem.set_param("memory_type", 'Lambda895')} several characteristics of the memory, such as wavelength, polarization, bandwidth, lifetime are set automatically. This is important as the properties of the state that is to be stored, must match those of the memory.  The parameters used in the memory modeling $t_{\mathrm{in}}, t_{\mathrm{out}}, \kappa$ and $n_B$  are calculated from the  $\eta_{\mathrm{int}} $ and $\eta_{\mathrm{e2e}}$, as given in Table \ref{tab:memoryperformances}. We assume the read-in and read-out efficiency to be the same, such that $t_{in}= t_{out} =1-\sqrt{\eta_{\mathrm{int}}} $.
Alternatively, the users may specify their own parameters by setting the \lstinline{memory_type} parameter to \lstinline{Test}, and setting the operating parameters $t_{\mathrm{in}}, t_{\mathrm{out}}, \kappa_e, \kappa_l, n_{\mathrm{B},e}, n_{\mathrm{B},l}$ to custom values. The default values for the other parameters are $\lambda = 895\ \mathrm{nm} $,  $\tau_s = 1\ \mathrm{\mu s} $, $\Delta \omega = 500\ \mathrm{MHz}$, accepted polarization is either $H,V$ and the $\tau_{\mathrm{retrig}} = 1\ \mathrm{\mu s}$. 
To determine the correct operating efficiency, the storage time must be initialized. This is the duration between the read-in and read-out of the memory, $\tau_s$. As mentioned in Section \ref{sec:characteristics}, we consider only homogeneous decoherence effects and thus we model $\eta_\mathrm{int}(t)$, the internal storage efficiency at time $t$, as an exponential $\eta_{\mathrm{int}}(t) = \eta_{\mathrm{int}}(0) \exp(-t/\tau)$. The $1/e$ storage times, $\tau$ for each memory can be found in Table \ref{tab:memoryperformances}. Setting the storage time has two effects, it sets the internal efficiency of the memory according to the exponential decay above, and sets the operation time of the system. The user must take caution that this exponential decay does indeed hold for the memory they would like to use.
In the case where the transmisivities of the read-in and read-out beamsplitter  $t_{\mathrm{in}}, t_{out}$ are user-defined, setting the storage time does not vary the beamsplitter transmissivity, so varying the storage time will have no effect. In this case, it is left to the user to account for the decay in storage efficiency.
As the memory has its own internal state, the photon number at which the state is truncated must be set accordingly. This is done by setting the \lstinline{truncation} parameter to an integer.

\subsection{State initialization}
States that are to be stored in the memory, must have properties that match those of the memory. To ensure compatibility between the source and memory, several checks on the properties of incoming state are performed. Often, the relationship between the input photon properties and the effect on the memory operation is complex and has not been experimentally measured in all cases. As a result, we simplify the temporal and frequency behavior of the memory. A description of the source-memory compatibility checks and their limitations is provided below.  The code snippet \ref{lst:State_init} shows the initialization of a state compatible with the memory initialized in code snippet \ref{lst:Initalization}.

\begin{listing}
\begin{minted}{python}
path_a_prop = StateProp(state_type="light",
                        truncation=5,
                        wavelength=894,
                        polarization="V",
                        uuid="a",
                        bandwidth = 0.5e9)
\end{minted}    
\caption{The initialization of a state that is compatible with Lambda Cs D1 memory example above.}
\label{lst:State_init}
\end{listing}
The state to be stored in the memory has a specific wavelength and bandwidth, which should match that of the optimal operating conditions of the memory. Therefore, we apply a check that the wavelength of the state matches that of the memory within $1\ \mathrm{nm}$, $\lambda_{\mathrm{mem}} = \lambda_{\mathrm{state}}$. Since the frequency dependence of noise and internal efficiency has not been reported for all memory experiments included in this work, proper consideration of the frequency compatibility of source and memory should be carried out independently, and was deemed beyond the scope of this work. To determine the eligibility of pairing a single photon source with a memory, further investigation into wavelength compatibility is required.  Moreover, in the current implementation, photons with a bandwidth larger than that of the memory are rejected, $\Delta \nu_{\mathrm{mem}}\ \geq  \ \Delta \nu_{\mathrm{state}}$. Experimentally, the memory functions as a frequency filter, storing photons whose frequency falls within the memory bandwidth window. Thus, the bandwidth behavior of the simulation does not fully represent the experimental behavior of the memory, but remains a topic for further extension of this simulation.

In this work, we reduce the time continuity of a memory experiment to discrete-time Fock states, representing early and late time bins.  We explicitly do not consider the temporal shape and timing of the photons involved in the storage process. However, the photon envelope shape after storage in an ensemble-based vapor memory strongly depends on the temporal waveform shape of the control pulses. \cite{novikova_optimal_2007}. Furthermore, the read-in and read-out processes from the vapor memory may result in dispersion or phase distortion of the photon wavepacket. The extent of these effects is not fully understood, and remains a topic of ongoing research.

Finally, the polarization of the input state is compared to that of the memory. If they are in the same polarization orientation, the state is stored.   
We assume that the average noise photon $\mu_1$ remains the same for both polarization orientations of the memory.

%Memories only accept a specific polarization orientation. We allow the user to select between two different directions of a given polarization orientation. Note that we specify the polarization measured. This is relevant for the suppression of noise as some systems use polarization suppression optics to reduce noise contributions. Consequently, the single to noise ratio can only be assured for polarization in the same direction as measured by the authors. 

\subsection{Memory channel query}

After initialization, the memory is queried, and the Kraus operators representing the read-in and read-out operations are returned and applied to the corresponding states. An example of a read-in operation is shown in code snippet \ref{lst:mem_query}.
\begin{listing}
\begin{minted}[fontsize=\scriptsize]{python}
response, operators = mem1.channel_query(whole_state, {"input"  : state_path_a_h_prop.uuid, 
                                                       "op_type": 'storage'})
whole_state.apply_kraus_operators(operators, whole_state.get_all_props(response["kraus_state_indices"]))
\end{minted}    
\caption{An example query of the memory, returning the Kraus operators for the read-in operation. Here the type of operation ('storage', 'retrieval) is chosen by setting 'op\_type': 'storage'. The read-in operation is then applied to the input state, and the internal state of the memory }
\label{lst:mem_query}
\end{listing}
%[caption={A  }, label={lst:mem_query}, basicstyle=\fontsize{8}{10}\selectfont\ttfamily]
To represent the different operation timescales of the memory, a memory query returns two different timings: the operation time and the re-triggering rate. The operation time is the storage time specified by the user in initialization. The re-triggering time is the time until a memory experiment can be repeated. Experimentally, this is the operation time plus the time taken for other experimental processes, such as pumping, cooling, and trigger delays.  If no pumping is present in the experiment, the re-triggering rate is assumed to be the operation time. The operation and re-triggering times are given in Table \ref{tab:memory_triggertimes}. 
When a photon is stored in the memory, that is when the channel is queried with op\_type = storage, re-trigger is set to false. Upon a channel query with the op\_type = retrieval, the re-trigger is set to true.  We assume that the passage of a second single photon through the memory during the storage time does not result in further decoherence of the memory state. 

\section{Examples}\label{sec:examples}

In the following section, we present example applications of digital twins, specifically the simulation of a Mach-Zehnder interferometer with a memory element in one of its arms, and the storage of quantum tokens. 

\subsection{Visibility and fidelity simulations for memories}
\begin{figure*}[h]
    \centering
    \subfloat{\includegraphics[width=0.9\textwidth]{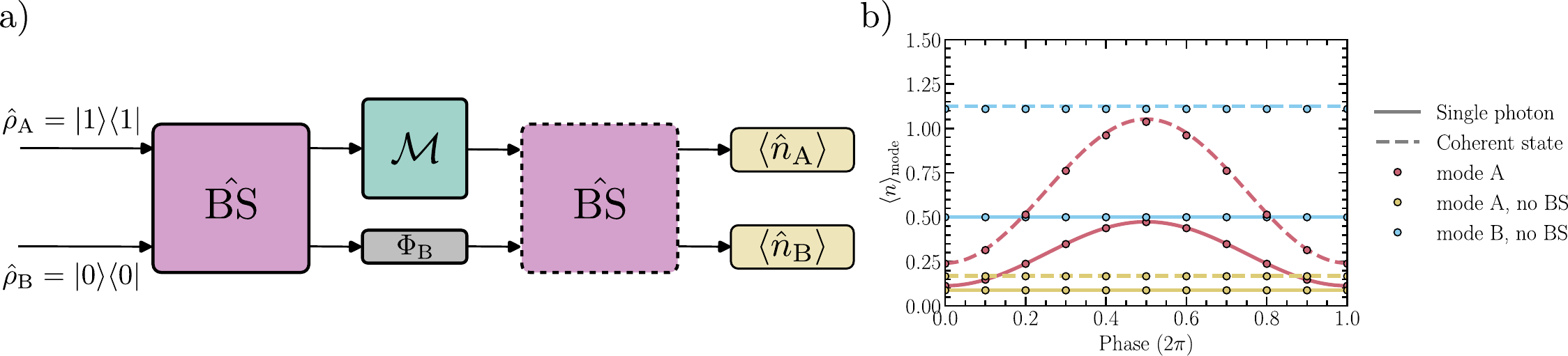}}
 
    \caption{a) Quantum channel diagram for a Mach-Zehnder Interferometer (MZI). A single photon is sent into path mode A and the average photon numbers in mode A and mode B are measured. b) The average photon number $\langle \hat{n} \rangle$ in mode A of the MZI is shown in red, for a single photon input (solid line) and for a coherent state input $\alpha = 1.5$ (dashed line). The blue and yellow curves show the average photon number in modes A and B, respectively, with the second beamsplitter removed. The data points represent simulated values, while the curves correspond to theoretical predictions from analytical solutions (see MZI section of the supplementary document). An excellent agreement between the analytical predictions and simulated results is observed. The truncation for the coherent state simulation is $n =7$. }
    \label{fig:MZI_overview}
\end{figure*}

\begin{figure*}[h]
    \centering
    \qquad
  \includegraphics[width=0.8\textwidth]{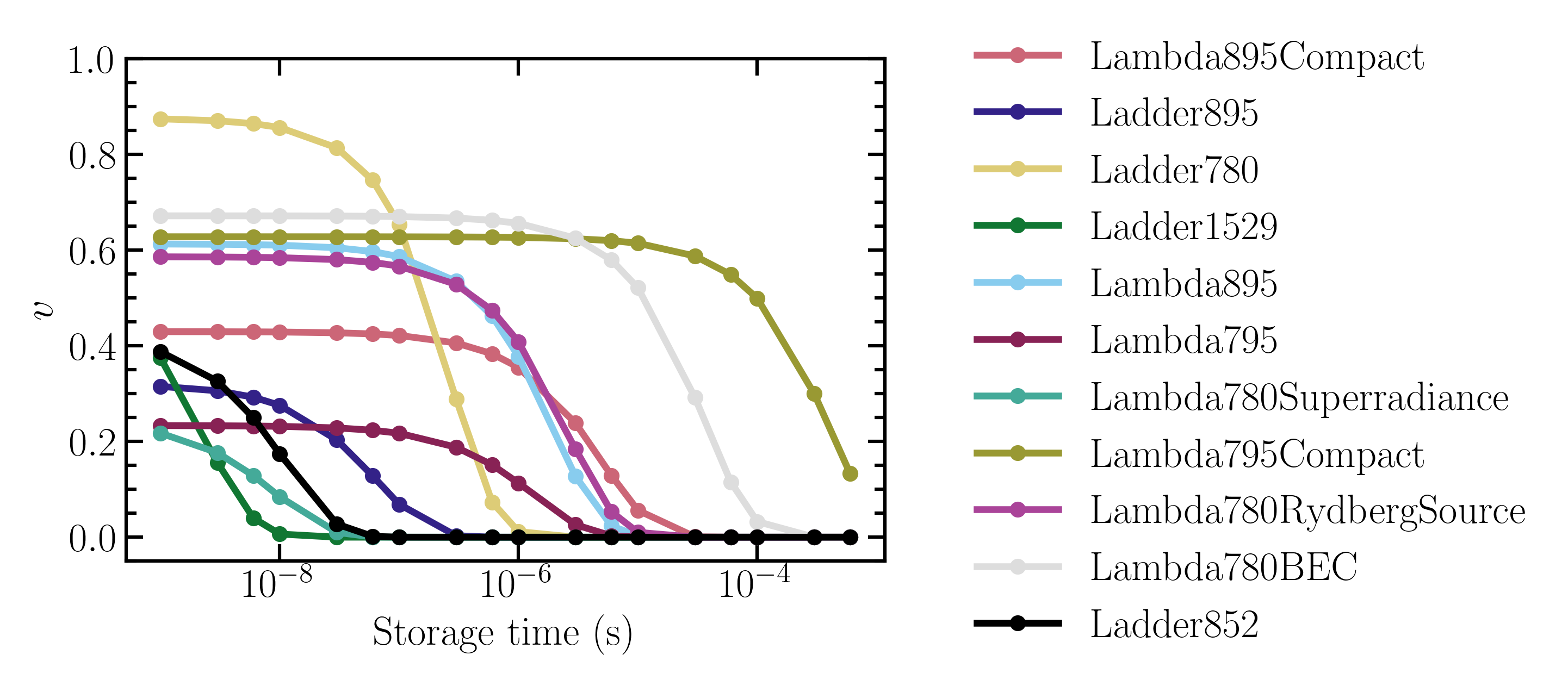}
    \caption{ Visibility of the single photon MZI interference fringes after storage in a memory, as a function of storage time.}
    \label{fig:MZI_comparison}
\end{figure*}

To demonstrate the compatibility of the memory simulation with both single photons and coherent states,  we measure the interference fringes observed when states are sent into a Mach-Zehnder Interferometer (MZI) with a phase element in one arm, and a memory element in the other (as shown in Figure \ref{fig:MZI_overview}a). The beamsplitters are assumed to have 50\% transmission/reflection and the memory implementation in \cite{esguerra_optimization_2023} has been chosen as an example, due to its high noise,  $\mu_1 = 0.07$. The path interference for a single photon input state and coherent state input with $\alpha = 1.5$ is shown in Figure \ref{fig:MZI_overview}b. We derive analytical solutions for the average photon number in mode A, for single photons and coherent states, which are depicted as full and dashed lines respectively (see supplementary information). As one can see, we observe excellent agreement between simulation and theory. To examine the behavior of the system without interference between the two modes, we remove the second beamsplitter. The memory exhibits a zero-time end-to-end efficiency of $\eta_\mathrm{e2e}(0)=0.13$ which reflects in the lowered transmission through path A in the absence of the second beamsplitter (plotted in yellow). The visibility in an MZI configuration is an important measure for the indistinguishability of quantum states which has to be preserved by quantum memories in practical applications. Indeed for single photons, fringe visibility is an upper bound on the two photon interference visibility value of photons stored in a quantum memory. We calculate the visibility for a mode, given as: 

$$v = \frac{\langle \hat{n}\rangle _{\mathrm{max}}-\langle \hat{n} \rangle_{\mathrm{min}}}{\langle \hat{n} \rangle_{\mathrm{max}}+\langle \hat{n} \rangle_{\mathrm{min}}}.$$
Here $\braket{\hat{n}}_{\mathrm{max}} (\braket{\hat{n}}_{\mathrm{min}})$ are the maximum (minimum) average number of photons measured in a single arm. We see in Figure \ref{fig:MZI_overview} b) that the presence of noise from the memory lowers the visibility of the interference.
 In Figure \ref{fig:MZI_comparison} we compare the fringe visibility in the MZI configuration for some of the most recent atomic vapor memories (see Table \ref{tab:memoryperformances}) assuming single-photon input. We find that memories with both high efficiency and lowest noise figure exhibit the highest fringe visibility. As expected, the visibility for all memories decreases exponentially with time according to their decrease in efficiency.

\subsection{Quantum token protocol}

To demonstrate an application of the memory simulation, we simulate a quantum token protocol, based on the experimental implementation described in \cite{bozzio_experimental_2018}. Quantum token protocols are used to safely encode classical information in qubits using either one of two basis sets $x$ and $z$, i.e. $x = \{\ket{0},\ket{1}\}, z=\{\ket{+},\ket{-}\}$. Due to the no-cloning theorem, only a party knowing the basis used to encode the classical bit can unambiguously recover the encoded information. Imperfections in such protocols are described in terms of correctness parameters $c_{ij}$, which is the probability of recovering the correct classical bit that was encoded in basis $i=x,z$ when measuring in basis $j=x,z$. Here we focus on the composite correctness parameter $c = \frac{1}{2}(c_{xx}+c_{zz}$, which yields $c=1$ for a perfect quantum token experiment and $c=1/2$ for a token experiment exhibiting random behaviour.

\begin{figure}[h]
    \centering
    \includegraphics[width=1\linewidth]{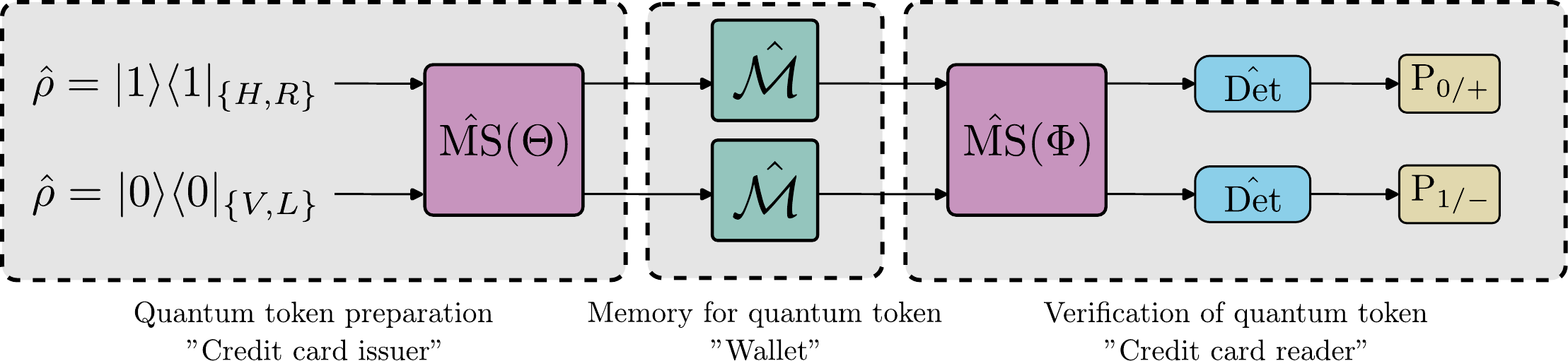}
    \caption{Quantum channel description of a quantum token protocol. A polarization encoded quantum token is prepared (either in H and V or L and R) and stored in two memories, one for each orthogonal polarization. After the retrieval from the memory the token is decoded and the click probability is measured on noisy orthogonal detectors.}
    \label{fig:token_schematic}
\end{figure}

In the experimental implementation demonstrated in \cite{bozzio_experimental_2018} the computational basis states are mapped onto polarization states of the photons. Given the memories accept a single polarization orientation, the basis must match that of the memory. Thus for memories which accept $H/V$ polarization orientations we map $\ket{0} = \ket{H} , \ket{1} = \ket{V}$, and $\ket{+} = \ket{D}, \ket{-} = \ket{A}$.  For the $R/L$ polarization basis $\ket{0} = \ket{R} , \ket{1} = \ket{L}$, and $\ket{+} = \ket{H}, \ket{-} = \ket{V}$.  In the model, polarization is given as labeled Fock states, with orthogonal polarizations having two different states, e.g  $\ket{+} = \frac{1}{\sqrt{2}}(\ket{1}_H + \ket{1})_V$.  To prepare a token in a given state, a horizontal Fock state, and the vacuum vertical state are sent to the tunable mode selector ($\hat{MS}$). The mode selector acts as a beam splitter with the matrix representation: 

\begin{equation}
    \hat{MS}(\theta) = \begin{pmatrix}
\cos2\theta & \sin2\theta  \\
\sin2\theta & -\cos2\theta 
\end{pmatrix} 	.
\end{equation}
By setting the $\theta  = \pi/2,\ 3\pi/4 , 3\pi/8, 5\pi/8$  the incoming $\ket{1}_{\{H,R\}} + \ket{0}_{\{V,L\}}$, returns the states $\ket{0},\ \ket{1}, \ \ket{-} ,\ \ket{+}$, respectively.  In the $H/V$ state encoding basis, the mode selector can be considered as the combined perfect action of a $\lambda/2 $ waveplate and a polarizing beamsplitter.  After storage in two memories that accept the orthogonal polarization directions,  the states are then read out of the memories and sent to a second mode selector, which sets the measurement basis. Setting $\Phi  = \pi/2\ ( 5\pi/8)$, sets the measurement basis to the $z (x)$  basis, respectively.  We model detector loss and dark counts using the Kraus operators presented in \ref{sec: Loss and Noise}, where $\kappa = 0.25$ and $n_{b} = \frac{7\times 10^{-5}}{1 - \kappa}$, corresponding to the values given in \cite{bozzio_experimental_2018}.

 We consider the calculation of $c_{zz}$ from the simulation in Figure \ref{fig:token_schematic}. The correctness of the measurement of photons prepared in the $z$ basis, is an average of measuring the $\ket{0} $ and $\ket{1}$ states correctly, $c_{zz} = \frac{1}{2 }(c_0 + c_1)$.  The preparation of perfectly polarized photons in state $\ket{0}$ and measurement in the $z$ basis, will result in a click in the detector in path mode $0/+$, measuring '0'. Likewise, the preparation of a photon in state $\ket{1} $ will result in a click in the detector in path  $1/-$, measuring the state to be  '1'. Sending in the state $\ket0$ and measuring a click on detector $1/-$, corresponds to a wrong result, and thus reduces the correctness.  The correctness of measuring $\ket{0}$, given a measurement in the z basis is,  $c_0 = \frac{P_{\mathrm{click\ only \ 0
 /+ }}}{ P_{\mathrm{click\ either}}}$ , where $ P_{\mathrm{click\ only \ 0/+}}$ is the probability of getting a click only in the detector in path measuring the states $\ket{0},\ \ket{+}$ and $P_{\mathrm{click\ either}}$ is the probability of getting at least one click in either arm.  Thus if $P_{0/+}$ is the probability of measuring a photon as $0/+$, and $P_{1/-}$ the probability of measuring a photon as being in state $1/-$, then $c_0 = \frac{P_{0/+}  (1-P_{1/-})}{1-(1-P_{0/+})(1-P_{1/-}
)}$and similarly the correctness of measuring $\ket{1}$, $c_1 =  \frac{ P_{1/-}  (1-P_{0/+})}{1-(1-P_{1/-})(1-P_{0/+})}$. The probability of measuring at least one photon, is given by $P_{0/+}= \sum_{i = 1}^N\sum_{j = 0}^N \bra{i,j}\rho\ket{i,j}$,  $P_{1/-}= \sum_{i = 1}^N\sum_{j = 0
}^N \bra{j,i}\rho\ket{j,i}$, where the state $\ket{i,j} = \ket{i}_{\{H,R\}} \ket{j}_{\{V,L\}}$. To measure $c_{xx} = \frac{1}{2 }(c_+ + c_-)$, one repeats the simulation, with input states corresponding to $\ket{+}, \ket{-}$, and measurement in the $x$ basis.

%\begin{figure*}[h]
%    \centering
%    \subfloat{\includegraphics[width=0.90\textwidth]%{figures/memory_comparison_new.png}}
  %  \qquad
%    \subfloat{\includegraphics[width=0.76\textwidth]{figures/memory_comparison_storage_time_new.png}}
%    \caption{
%\end{figure*}

\begin{figure}
    \centering
    \includegraphics[width=\linewidth]{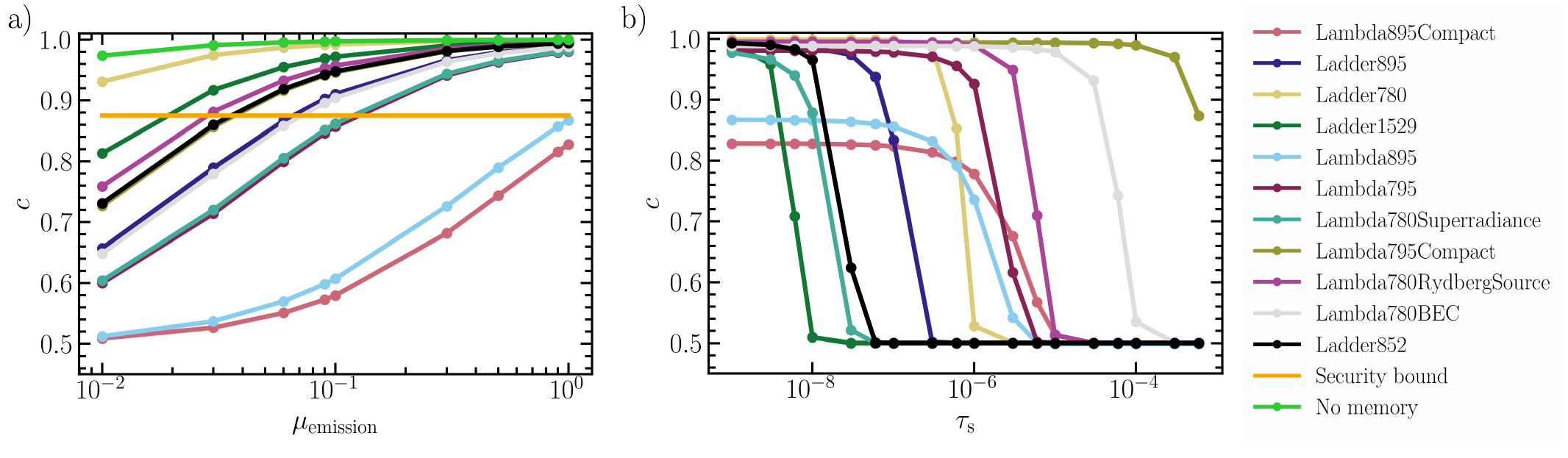}
    \caption{a) Simulated correctness values for different memories used in the protocol, with $\tau_s = 0 $. The expected correctness of the system without memory is shown in green.  b) The variation of the correctness of a memory token as a function of storage time for $\mu_{\mathrm{emission}} = 1$. 'Lambda795Compact' has a lifetime of $180\ \mu \mathrm{s}$, enabling high token correctness for long storage times. }
    \label{fig:Quantum_token_comparison}
\end{figure}

Figure \ref{fig:Quantum_token_comparison} a) shows the correctness of quantum tokens stored in different memories, with respect to the single photon emission probability, $\mu_{\mathrm{emission}}$.  For perfect on-demand photons, $\mu_{\mathrm{emission}} = 1$, most memory implementations achieve a correctness above the single photon correctness threshold to ensure the security of the quantum token protocol, $c > 7/8 $ \cite{bozzio_experimental_2018} (shown in orange). However, achieving an almost perfectly efficient single photon source with high purity is experimentally challenging. A clear factor influencing the correctness for low photon emission probabilities is the signal to noise ratio. Memories based on the ORCA/FLAME protocol with SNR on the order of $\sim 10^{3}$, result in the highest correctness. The two memories with the highest $\mu_1$ do not reach the security threshold, even for a perfect single-photon emission probability.  Importantly, these results give upper bounds for the correctness, as $\tau_s = 0$.  To simulate the correctness of different memories at varying storage times, we set $\mu_{\mathrm{emission}}=1$, and scan $\tau_s$ between $1\ \mathrm{ns}$ to $600\ \mathrm{us}$. Figure \ref{fig:Quantum_token_comparison}
b) shows the storage time dependence of the correctness parameter of each memory implementation.

\section{Conclusion}\label{sec:conclusion}

We develop a quantum channel description of ensemble based atomic vapor memories and provide model parameters for state-of-the-art quantum memory experiments. We use our model to estimate upper bounds for the visibility of a HOM-type measurement for single photons stored in a memory and observe excellent agreement between theory and simulations. Furthermore, we use the digital twin of the memories in assessing the security of a quantum token protocol. We compare the performance of different memory implementations from the literature. To further extend the applicability of the digital twins within quantum network simulations, proper consideration of frequency and temporal effects is needed. Matching of the memory bandwidth and photon temporal wavepacket as well as the inclusion of dispersive effects is subject of future work. The model presented allows for easy extension to include and compare other optically controlled memories such as solid-state memories. The presented digital twins are embedded in an easy-to-use simulation framework which allows for straightforward estimates of memory performances in multi-device quantum networks and are a step toward full end-to-end experimental simulation of quantum optics experiments.

\subsubsection*{Acknowledgments}
We would like to thank Leon Messner for humorously insightful discussions. The authors acknowledge financial support from DFG project 448532670 and BMFTR through project 16KIS1717K.

\subsubsection*{Data availability}
The data is available from the authors upon reasonable request.

\subsubsection*{Conflict of interest}

The authors declare no potential conflict of interests.

\subsubsection*{Supporting information}

Additional supporting information may be found in the
online version of the article at the publisher’s website.
\bibliographystyle{wileyNJD-APS}
\bibliography{references.bib}
%\bibliography{Digital_twin_memory}

\newpage
\appendix

\begin{sidewaystable}[h!]%
  \begin{threeparttable}

\caption{This table collects the re-trigger time, polarization acceptance and memory protocols of some of the most recent atomic vapor memories.\label{tab:memory_triggertimes}}
\begin{tabular*}{\textwidth}{@{\extracolsep\fill}lllll@{}}
\toprule
\textbf{Reference} & \textbf{Polarization}  & \textbf{re-trigger time} & \textbf{memory protocol} & Remarks \\
\midrule
 Jutisz et al. (2025) \cite{jutisz_stand-alone_2025}& linear& 32.7\,$\mu$s  & $\lambda$-configuration (Raman EIT)&coherent states at single-photon level\\

  Maa\ss\, et al. (2024) \cite{maas_room-temperature_2024}& linear (V measured) & 33\,ns&ladder-configuration (FLAME)&coherent states at single-photon level\\

Davidson et al. (2023) \cite{davidson_fast_2023} & circular ($\sigma^+$ measured)  & 108\,ns & ladder-configuration (FLAME)&coherent states at single-photon level\\

 Thomas et al. (2023) \cite{thomas_single-photon-compatible_2023} &linear & 12.5\,ns & ladder-configuration (ORCA)&coherent states at single-photon level\\

  Esguerra et al. (2023) \cite{esguerra_optimization_2023}&linear& 11\,$\mu$s& $\lambda$-configuration (Raman EIT)&coherent states at single-photon level\\

 Buser et al. (2022) \cite{buser_single-photon_2022}& circular ($\sigma^-$ measured) & 2.7\,$\mu$s & $\lambda$-configuration (Raman EIT)& SPDC single photons\\
 
 Rastogi et al. (2022) \cite{rastogi_superradiance-mediated_2022}&linear&5.7\,$\mu$s $^{\rm *}$ & $\lambda$-configuration (SR-mediated)&coherent states at single-photon level\\
 
  Wang, Craddock et al. (2022) \cite{wang_field-deployable_2022} &circular ($\sigma^-$ measured)&5\,ms& $\lambda$-configuration (Raman EIT)&coherent states at single-photon level\\

   Heller et al. (2022) \cite{heller_raman_2022}& circular & 11ms & $\lambda$-configuration (Raman EIT)& Lambda780RydbergSource\\

   Saglamyurek et al. (2022) \cite{saglamyurek_storing_2021}& circular & 20s & $\lambda$-configuration (ATS)\\

   Kaczmarkek et al. (2018) \cite{kaczmarek_high-speed_2018}& linear (H measured) & 12.5\,ns & ladder-configuration (ORCA)&SPDC single photons \\
   
 \midrule

   Wei et al. (2020) \cite{wei_broadband_2020} & circular ($\sigma^+$ measured) & n.a. & $\lambda$-configuration (Raman EIT)&coherent state at single-photon level\\
   
 Vernaz-Gris et al. (2018) \cite{vernaz-gris_highly-efficient_2018}&circular ($\sigma^+$ measured)&50\,ms & $\lambda$-configuration (Raman EIT)&coherent states at single-photon level\\
   
  Katz et al. (2018) \cite{katz_light_2018}& linear (H measured) & n.a. & $\lambda$-configuration (Raman EIT)&coherent states\\

\bottomrule
\end{tabular*}
\begin{tablenotes}%%[341pt]
\item[$^{\rm *}$] This value does not include the preparation time of the cold ensemble which is on the order of ms.

\end{tablenotes}
\end{threeparttable}
\end{sidewaystable}

\begin{sidewaystable}[h!]%
  \begin{threeparttable}

\caption{This table collects the performance metrics of some of the most recent atomic vapor memories.\label{tab:memoryperformances}}
\begin{tabular*}{\textwidth}{@{\extracolsep\fill}llllllllll@{}}
\toprule
\textbf{Reference} & \textbf{Atomic  species}  & \textbf{$\lambda$ [nm]}  & {\textbf{$\eta_{\mathrm{e2e}}$ ($\tau_\mathrm{s}=0$)}}  & $\mathbf{\mu}_1$ & \textbf{Bandwidth} & \textbf{1/e Storage time} &  Class name \\
\midrule

 Jutisz et al. (2025)$^{\rm a}$ \cite{jutisz_stand-alone_2025}&Cs& 895&0.054 (0.23 internal)&$0.06$ &>\,44\,MHz&2.4$\mu$s&Lambda895Compact\\

 Maa\ss\, et al. (2024) $^{\rm a}$ \cite{maas_room-temperature_2024}& Cs& 895& (0.027 (0.210 internal) &$7.2\times10^{-5}$ &560\,MHz&32\,ns&Ladder895 \\

Davidson et al. (2023) $^{\rm a}$ \cite{davidson_fast_2023} &Rb  & 780 & 0.35 (0.51 internal)  & $3\times 10 ^{-6}$&370\,MHz&108\,ns& Ladder780 \\

 Thomas et al. (2023)$^{\rm a}$ \cite{thomas_single-photon-compatible_2023} &Rb&1529&0.11 (0.21 internal)& $4.4\times10^{-6}$ &>\,1\,GHz&1.1\,ns&Ladder1529\\

  Esguerra et al. (2023)$^{\rm a}$ \cite{esguerra_optimization_2023}&Cs&895& 0.13 (0.33 internal) & 0.07 & 220\,MHz&140\,ns& Lambda895 \\

 Buser et al. (2022) $^{\rm a}$ \cite{buser_single-photon_2022}& Rb & 795& 0.014 (0.047 internal)   & $2.4\times10^{-5}$   &370\,MHz&680\,ns& Lambda795 \\

 Rastogi et al. (2022)$^{\rm b}$ \cite{rastogi_superradiance-mediated_2022}&Rb&780&0.015$^{\rm c}$ (0.03 internal)&$2.1\times10^{-4}$ &12.7\,MHz&4.7\,$\mu$s& Lambda780Superradiance \\
 
  Wang, Craddock et al. (2022)$^{\rm a}$ \cite{wang_field-deployable_2022}&Rb&795&0.125$^{\rm c}$ (0.25 internal)&$1.9\times10^{-3}$ &2\,MHz&180\,$\mu$s&Lambda795Compact\\

   Heller et al. (2022) $^{\rm b}$ \cite{heller_raman_2022}& Rb&780&0.105$^{\rm c}$ (0.21 internal)&$1.0\times10^{-3}$&17.6\,MHz&$1.2\,\mu$ s&Lambda780RydbergSource\\

   Saglanyurek et al. (2021) $^{\rm b}$ \cite{saglamyurek_storing_2021}& Rb & 780 &0.15$^{\rm c}$ (0.3 internal) & $5\times10^{-3} $&  22 MHz & 15.8$\mu$s& Lambda780BEC\\

   Kaczmarkek et al. (2018) $^{\rm a}$ \cite{kaczmarek_high-speed_2018}& Cs  & 852  & 0.049 (0.17 internal)  &$ 3.8\times 10^{-5} $&1000\,MHz&5.4\,ns &Ladder852\\

 \midrule
  Wei et al.(2020) $^{\rm b}$ \cite{wei_broadband_2020}  & Cs & 895& 0.8 internal& n.a.&31.4\,MHz&54$\mu$s\\
 
 Vernaz-Gris et al. (2018)$^{\rm b}$ \cite{vernaz-gris_highly-efficient_2018}&Cs&852& 0.70 internal&n.a.& 2\,MHz&14\,$\mu$s&\\

  Katz et al. (2018)$^{\rm a}$ \cite{katz_light_2018}& Cs & 895&  0.09 internal& n.a. &n.a.& 149\,ms & \\

\bottomrule
\end{tabular*}
\begin{tablenotes}%%[341pt]
\item[$^{\rm a}$] Operated at room-temperature.
\item[$^{\rm b}$] Operated at cryogenic temperature.
\item[$^{\rm c}$] No value provided, calculated assuming $\eta_{\mathrm{transm}} = 0.5$.

\item {\it Remark:} We do not provide fitting errors for the value of n$_\mathrm{b}$ because they are much smaller than the measurement errors of the SNR, which is not always provided in the literature. The given end-to-end efficiencies correspond to the extrapolated "zero time" efficiencies as it is convention in the community. We want to emphasize that the practical meaning of this measure is questionable as all memories will have lower efficiencies for any practical storage time.
\end{tablenotes}
\end{threeparttable}

\end{sidewaystable}

\section{Truncation}\label{app:truncation}
\begin{figure}[h]
     \centering
     \includegraphics[width=0.5\linewidth]{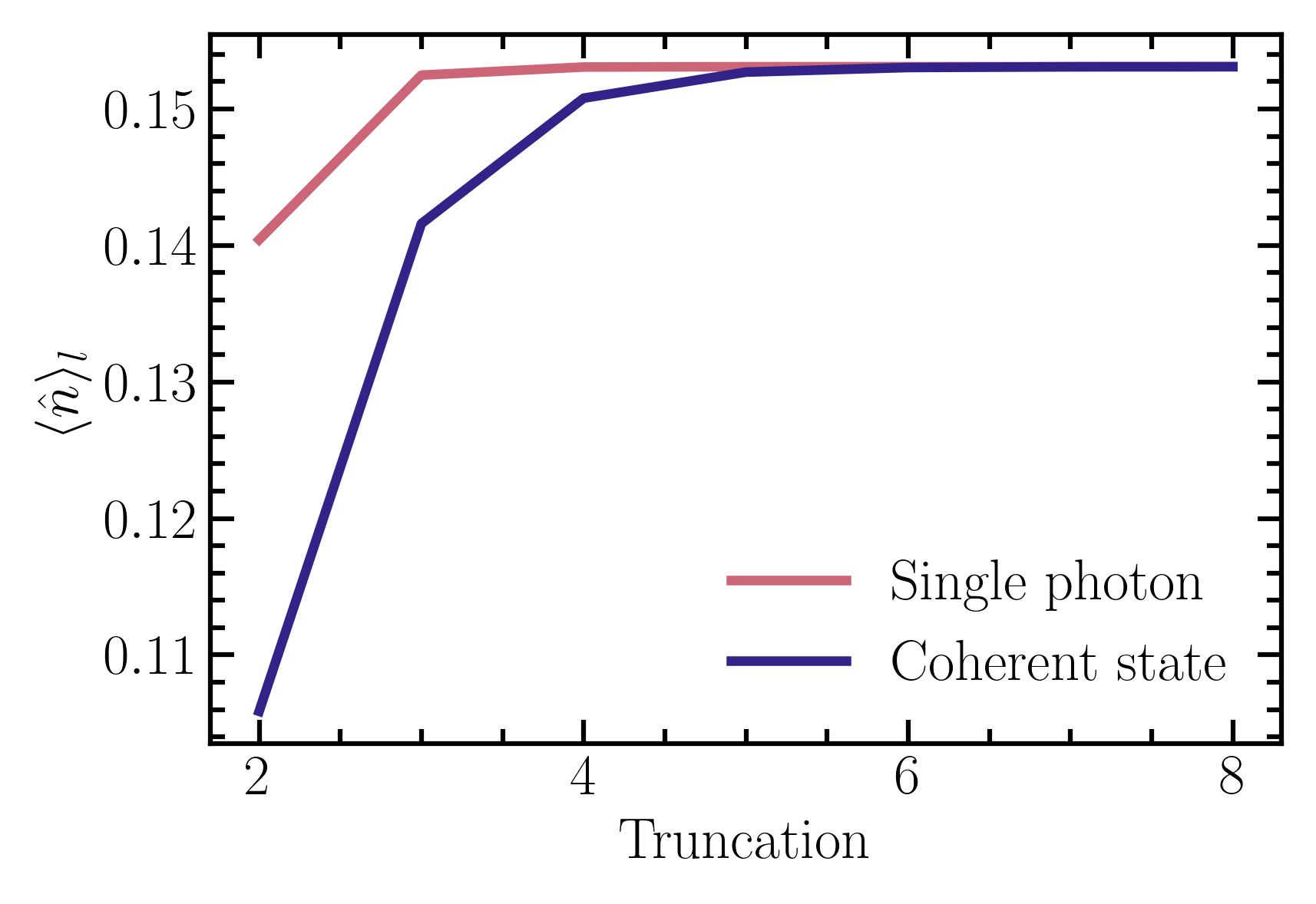}
     \caption{Convergence of the average photon number after storage in a 'Lambda895' memory. }
     \label{fig:truncation}
 \end{figure}

We analyze the truncation of our presented memory simulation by varying the truncation of a memory experiment for the memory with the lowest signal-to-noise ratio (highest $\mu_1$). We simulate a memory experiment with the input state $\hat{\rho} = \ket{1}\bra{1}$ and $\hat{\rho} = \ket{\alpha }\bra{\alpha}$ where $\alpha = 1$, respectively. We measure the average photon number of the late time-bin, and observe the convergence with increasing truncation. Figure \ref{fig:truncation} shows the convergence of the expected photon number at a truncation of 5 for coherent states, and 3 for single photons. To simulate coherent states with $\alpha >1 $ a higher truncation will be required. This quickly scales the size of the state space needed to simulate the memory, making simulation of coherent states impractical for $\alpha \gtrapprox 1 $.

\section{Fidelity simulations for test memories}\label{app:Fidelity}

When assessing the performance of a quantum memory, a central question is: To what extent is the quantum state input to the memory preserved after retrieval? A natural approach is to evaluate the state overlap between the retrieved state $\hat{\sigma}$, and the input state ${\hat{\rho}}$, defined as:  
\begin{equation}
    \mathcal{F}(\hat{\rho},\hat{\sigma})=\left(\mathrm{tr}\sqrt{\sqrt{\hat{\rho}}\hat{\sigma}\sqrt{\hat{\rho}}}
\right)^2.
\end{equation}

By construction $\mathcal{F} = 1$ if $\hat{\sigma} = \hat{\rho}$. A benefit of our simulation is that we can easily access both $\hat{\sigma}$ and $\hat{\rho}$ , which in experiment typically would require two full state tomography measurements. However, we show below that care must be taken when interpreting this quantity, especially for different input states and in the presence of loss and noise.

\begin{figure*}[h]
    \centering
     \includegraphics[width=0.9\linewidth]{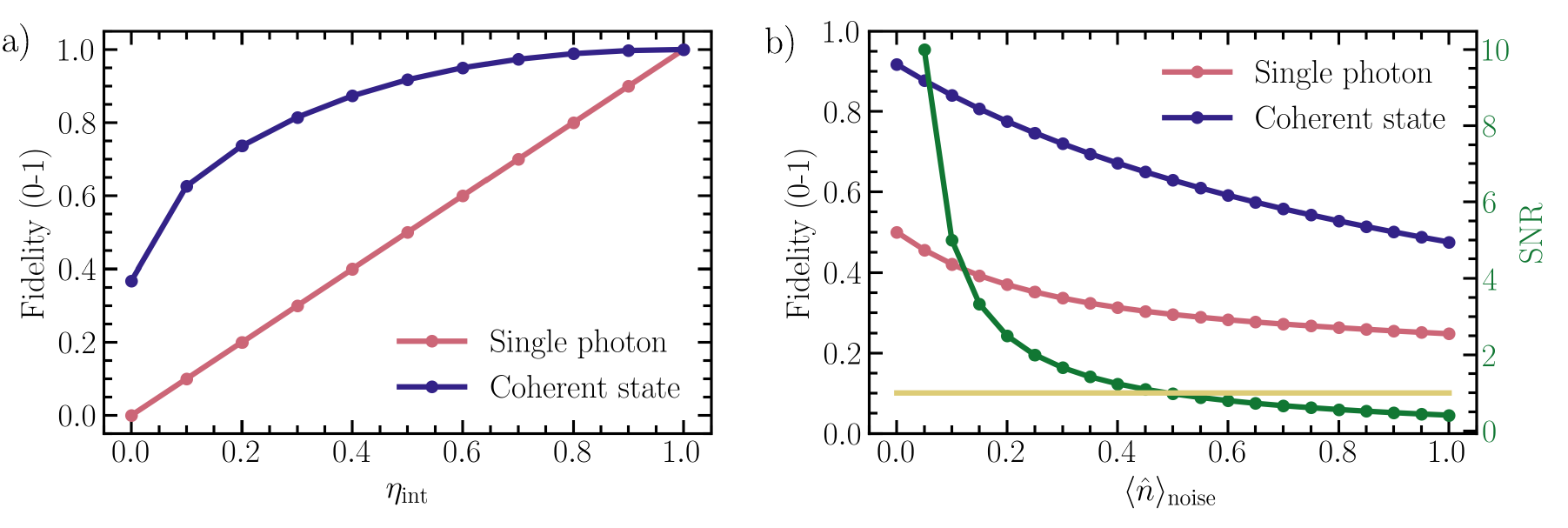}
    \caption{a) Variation of the fidelity of photons out of a memory, with no noise, and varying internal efficiency.  b) Fidelity of a memory with perfect internal efficiency, loss of 0.5, and varying noise photons. The signal-to-noise ratio is plotted in green, and the bound SNR = 1 is shown in yellow.}
    \label{fig:test_memory_fidelity}
\end{figure*}

The simulated fidelity values for two types of input states — single photons and coherent states with $\alpha=1$ —are illustrated in Figure \ref{fig:test_memory_fidelity}a. Here we simulate an ideal memory with perfect transmission efficiency ($\eta_{\mathrm{trans}}=1$) and no background noise ($n_{\mathrm{B}}=0$). For single photons, the fidelity increases linearly with the internal efficiency of the memory, reflecting the improving preservation of the input state. In contrast, the fidelity of coherent states does not vanish even as the internal efficiency $\eta_{\mathrm{int}}$ approaches zero. This is due to the non-zero vacuum component of coherent states — when the output is vacuum, the overlap with the original state still yields a fidelity of $\mathcal{F} = e^{-|\alpha|^2}$.

By measuring fidelity and signal-to-noise ratio (SNR)  with increasing noise photon number, we further observe the difference in fidelity between coherent states and single photon inputs, as seen in Figure \ref{fig:test_memory_fidelity} b). While SNR decreases with added noise --  as expected -— the fidelity of both states saturates to a non-zero value. 
Moreover, depending on the input state, the measured fidelity varies between input coherent state, and single photon - for the same SNR. Additionally, for the same SNR, the measured fidelity can differ depending on whether the input state is a single photon or a coherent state. This behavior underscores the importance of comparing memory fidelities only when the same input state is used, as differences in input can lead to misleading conclusions about performance.

%At low SNR, the fidelity becomes increasingly dominated by the overlap between noise photons and the input state, rather than reflecting the probability that the input state has been faithfully retrieved. Thus, fidelity in noisy regimes may overestimate memory performance by failing to penalize the presence of statistically similar but unrelated noise photons.

\begin{figure}
    \centering
    \includegraphics[width=0.8\linewidth]{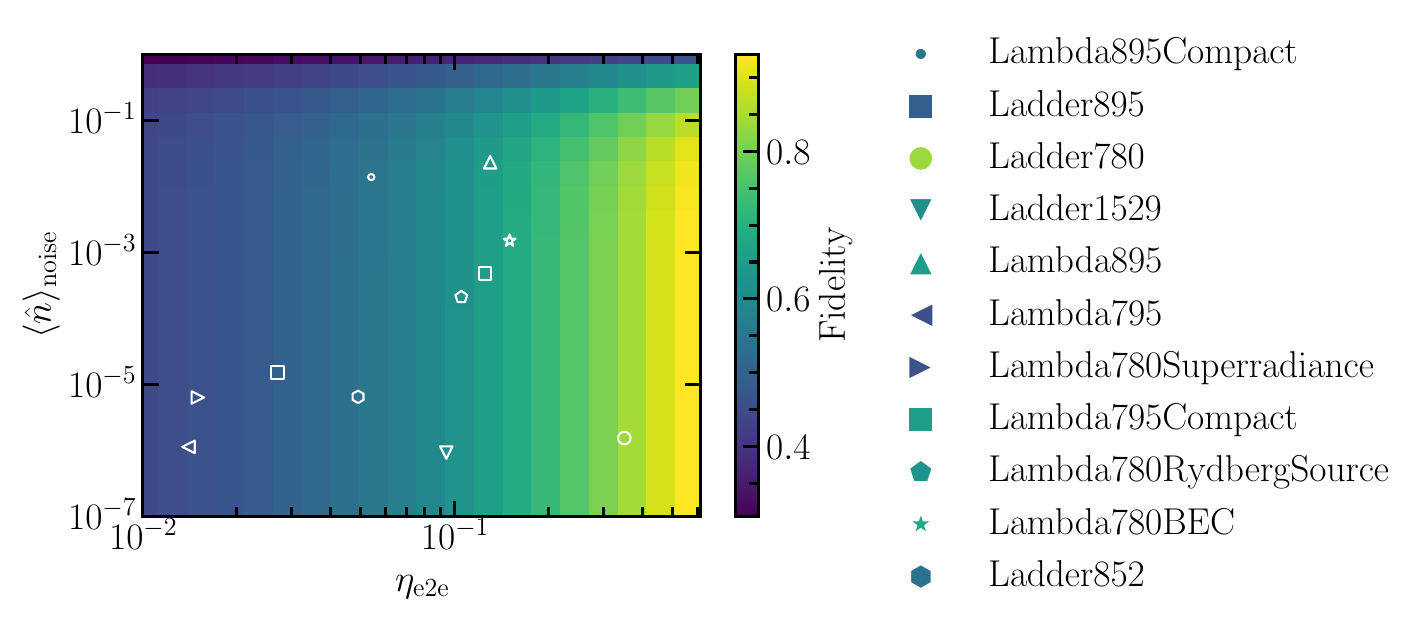}
    \caption{Simulated fidelity of different quantum memory experiments with and coherent state input $\alpha = 1$. The end-to-end efficiency has the highest influence on the fidelity of the retrieved state. In the experimentally achieved efficiency regime the number of noise photons doesn't influence the fidelity. Truncation $n = 5$.}
    \label{fig:fidelity_mems}
\end{figure}

Fidelity measurements for a coherent state $\alpha =1$ stored in various quantum memories are presented in Figure \ref{fig:fidelity_mems}. For high state fidelity after the memory, the improvement of the overall memory end-to-end efficiency is key.  

When assessing fidelity for single-photon inputs, it is important to recognize that the state $\hat{\rho} = \ket{1}\bra{1}$ has no phase coherence. As a result, the addition of incoherent noise photons can increase the fidelity -— not due to improved memory performance, but because the input state more closely resembles the incoherent noise. This effect should not be interpreted as an enhancement of memory quality.

In summary, fidelity remains a valuable figure of merit which we can easily measure in our model, however its interpretation requires careful consideration of the state input into the memory.  Specifically, fidelity comparisons should only be made between memories tested with the same input states, particularly in realistic, noisy, or lossy conditions.

%\usepackage[top=2cm,bottom=2cm,left=2cm,right=2cm,marginparwidth=1.75cm]{geometry}

%\usepackage{graphicx}
%\usepackage{amsmath,amstext,amssymb,amsfonts,graphics,graphicx,mathtools}
%\DeclareMathOperator{\Tr}{Tr}
%\usepackage{mleftright} % for '\mleft' and '\mright' macros
%\usepackage{times}
%\usepackage{caption}
%\usepackage{hyperref}
%\usepackage{titlesec}
%\usepackage{braket}
%\usepackage{gensymb}

% Change figure/table/equation/etc. numbering to start with "S"
\renewcommand{\thefigure}{S\arabic{figure}}
\renewcommand{\thetable}{S\arabic{table}}
\renewcommand{\theequation}{S\arabic{equation}}
\renewcommand{\thesection}{S\arabic{section}}

% Reset counters
\setcounter{figure}{0}
\setcounter{table}{0}
\setcounter{equation}{0}
\setcounter{section}{0}

\begin{center}
    {\Huge Supplemental material: A digital twin of atomic ensemble quantum memories}\\[1ex]  % Huge bold title
    {\large Elizabeth Jane Robertson, Benjamin Maa{\ss}, Konrad Tschernig, Janik Wolters}     % Optional smaller line
\end{center}

\section{Attenuating Beam Splitter Channel}\label{app:bs derivation}
In this section we want to derive a Kraus representation of the beam splitter interaction. First, we want to derive an operator representation $\hat{U}$ of  the beam splitter channel in the Heisenberg picture. The Schrödinger equation describing the beam splitter interaction is
\begin{equation}
	i \partial_z \ket{\psi} = \hat{H} \ket{\psi},
\end{equation}
with the Hamiltonian 
\begin{equation}
	\hat{H} = \omega \left( \hat{a}^\dagger \hat{a} + \hat{e}^\dagger \hat{e} \right) + \zeta \left(e^{i\phi}\hat{a}^\dagger \hat{e} + e^{-i\phi}\hat{a} \hat{e}^\dagger \right),
\end{equation}
where $\omega$ is the eigen-frequency of the modes, $\zeta$ is the coupling rate between them, $e^{i\phi}$ is a (fixed) relative phase that is picked up when the hoping occurs and $\hat{a}^\dagger (\hat{e}^\dagger)$ are the creation operators of the respective modes. This leads to the unitary evolution operator
\begin{equation}
	\hat{U}(z) = e^{-i z \omega  \left(\hat{a}^\dagger \hat{a} + \hat{e}^\dagger\hat{e}\right) - i z \zeta \left(e^{i\phi} \hat{a} \hat{e}^\dagger + e^{-i\phi} \hat{a}^\dagger \hat{e} \right) }.
	\label{eq:unitary_operator}
\end{equation}
Since, $\left[\hat{a}^\dagger \hat{a} + \hat{e}^\dagger\hat{e},  e^{i\phi}\hat{a} \hat{e}^\dagger +e^{-i\phi} \hat{a}^\dagger \hat{e} \right] = 0$ we can split the exponential
\begin{equation}
	\hat{U}(z) = e^{-i z \omega  \left(\hat{a}^\dagger \hat{a} + \hat{e}^\dagger\hat{e}\right) } e^{-i z \zeta \left( e^{i\phi}\hat{a} \hat{e}^\dagger + e^{-i\phi}\hat{a}^\dagger \hat{e} \right) }.
\end{equation}
The creation operator $\hat{a}^\dagger(0) = \hat{a}^\dagger$ evolves in the Heisenberg picture as
\begin{align}
	\hat{a}^\dagger(z) & = \hat{U}(z)\hat{a}^\dagger \hat{U}^\dagger(z) \\
	& =  e^{-i z \zeta \left( e^{i\phi}\hat{a} \hat{e}^\dagger + e^{-i\phi} \hat{a}^\dagger \hat{e} \right) } \underbrace{e^{-i z \omega  \left(\hat{a}^\dagger \hat{a} + \hat{e}^\dagger\hat{e}\right) } \hat{a}^\dagger e^{i z \omega  \left(\hat{a}^\dagger \hat{a} + \hat{e}^\dagger\hat{e}\right) }}_{=(*)} e^{i z \zeta \left( e^{i\phi}\hat{a} \hat{e}^\dagger + e^{-i\phi}\hat{a}^\dagger \hat{e} \right) } 
\end{align}
We first treat the $(*)$ term using the BCH formula $e^X Y e^{-X} = Y + [X,Y] +\frac{1}{2!} [X,[X,Y]] + \ldots $
\begin{align}
	(*) &= e^{-i z \omega  \left(\hat{a}^\dagger \hat{a} + \hat{e}^\dagger\hat{e}\right) } \hat{a}^\dagger e^{+i z \omega  \left(\hat{a}^\dagger \hat{a} + \hat{e}^\dagger\hat{e}\right) } \\
	&= \hat{a}^\dagger + \underbrace{\left[-i z \omega  \left(\hat{a}^\dagger \hat{a} + \hat{e}^\dagger\hat{e}\right),\hat{a}^\dagger \right]}_{= -iz\omega \hat{a}^\dagger}+\frac{1}{2!}\underbrace{\left[-i z \omega  \left(\hat{a}^\dagger \hat{a} + \hat{e}^\dagger\hat{e}\right),-iz\omega \hat{a}^\dagger \right]}_{=(-iz\omega)^2 \hat{a}^\dagger} + \ldots \\
	&= \hat{a}^\dagger e^{-iz\omega}.
\end{align}
Thus we have
\begin{align}
	\hat{a}^\dagger(z) & = e^{-iz\omega} e^{-i z \zeta \left( e^{i\phi}\hat{a} \hat{e}^\dagger + e^{-i\phi}\hat{a}^\dagger \hat{e} \right) }\hat{a}^\dagger e^{i z \zeta \left(e^{i\phi} \hat{a} \hat{e}^\dagger + e^{-i\phi}\hat{a}^\dagger \hat{e} \right) } \\
	& = e^{-iz\omega} \left(\hat{a}^\dagger + \underbrace{\left[-i z \zeta \left( e^{i\phi}\hat{a} \hat{e}^\dagger +e^{-i\phi} \hat{a}^\dagger \hat{e} \right),\hat{a}^\dagger  \right]}_{=-iz\zeta e^{i\phi} \hat{e}^\dagger } +\frac{1}{2!} \underbrace{\left[ -i z \zeta \left( e^{i\phi}\hat{a} \hat{e}^\dagger + e^{-i\phi}\hat{a}^\dagger \hat{e} \right), -iz\zeta e^{i\phi} \hat{e}^\dagger \right]}_{=(-iz\zeta)^2 \hat{a}^\dagger} + \ldots \right) \\
	&=  e^{-iz\omega} \left(\hat{a}^\dagger \underbrace{\left(1+\frac{(-iz\zeta)^2}{2!} +  \frac{(-iz\zeta)^4}{4!} +\ldots \right)}_{=\cos(\zeta z)} + e^{i \phi} \hat{e}^\dagger \underbrace{\left( -iz\zeta  +\frac{(-iz\zeta )^3}{3!} +\ldots \right)}_{=-i\sin(\zeta z)} \right) \\
	&= e^{-iz\omega} \left( \cos(\zeta z) \hat{a}^\dagger -i e^{i \phi} \sin(\zeta z) \hat{e}^\dagger \right).
\end{align}
In the same way we find 
\begin{equation}
	\hat{e}^\dagger(z) = e^{-iz\omega} \left( \cos(\zeta z) \hat{e}^\dagger -i e^{-i \phi} \sin(\zeta z) \hat{a}^\dagger \right).
\end{equation}
In summary, assuming the Hamiltonian $\hat{H} = \omega \left( \hat{a}^\dagger \hat{a} + \hat{e}^\dagger \hat{e} \right) + \zeta \left(e^{i\phi}\hat{a}^\dagger \hat{e} + e^{-i\phi}\hat{a} \hat{e}^\dagger \right)$ governs the beamsplitter dynamics, we obtain the transformed creation operators after time $z$
\begin{align}
	\hat{a}^\dagger(z) &= e^{-iz\omega} \left( \cos(\zeta z) \hat{a}^\dagger(0) -i e^{i \phi} \sin(\zeta z) \hat{e}^\dagger(0) \right) \\
	\hat{e}^\dagger(z) &= e^{-iz\omega} \left( \cos(\zeta z) \hat{e}^\dagger(0) -i e^{-i \phi} \sin(\zeta z) \hat{a}^\dagger(0) \right).
\end{align}
In order to match this transformation with the known beam splitter transformation:

\begin{equation}
	\hat{U}: \begin{pmatrix}
		\hat{a}^\dagger \\
		\hat{e}^\dagger
	\end{pmatrix} \rightarrow \begin{pmatrix}
	t & r \\
	-r & t
	\end{pmatrix} \begin{pmatrix}
	\hat{a}^\dagger \\
	\hat{e}^\dagger
	\end{pmatrix} = \begin{pmatrix}
	t \hat{a}^\dagger + r\hat{e}^\dagger \\
	-r \hat{a}^\dagger + t\hat{e}^\dagger 
	\end{pmatrix} ,
	\label{eq:unitary_rt}
\end{equation}

we choose $\omega = 0$, $z=1$, $\cos(\zeta) = t$, $\sin(\zeta)=r$, $\phi = \pi/2$. By plugging these values into Eq.~(\ref{eq:unitary_operator}) we thus found the operator representation of $\hat{U}$ 
\begin{equation}
	\hat{U} =  e^{\zeta \left( \hat{a} \hat{e}^\dagger - \hat{a}^\dagger \hat{e} \right) }.
\end{equation}

Second, we want to find the final state after the action of $\hat{U}$:
\begin{equation}
	\hat{P}^{(f)} = \hat{U}\hat{P}^{(i)} \hat{U}^\dagger = \hat{U}\hat{\rho}^{(i)} \otimes \hat{\rho}^{(e)}\hat{U}^\dagger.
\end{equation}
We now trace out the environment to find the reduced final system state
\begin{equation}
	\hat{\rho}^{(f)} = tr_e \left[ \hat{U}\hat{\rho}^{(i)} \otimes \hat{\rho}^{(e)} \hat{U}^\dagger \right].
\end{equation}
For the trace we choose the Fock basis and by rearranging the terms we find the Kraus operators
\begin{align}
	\hat{\rho}^{(f)} &= \sum_{l=0}^\infty \bra{l}_e \hat{U}\hat{\rho}^{(i)} \otimes \hat{\rho}^{(e)} \hat{U}^\dagger \ket{l}_e \\
	&=\sum_{l=0}^\infty \bra{l}_e \hat{U}\hat{\rho}^{(i)} \otimes   \ket{0}\bra{0} \hat{U}^\dagger \ket{l}_e \\
	&=\sum_{l=0}^\infty \underbrace{\bra{l}_e \hat{U} \ket{0}_e}_{\hat{W}_l} \hat{\rho}^{(i)} \underbrace{\bra{0}_e \hat{U}^\dagger \ket{l}_e}_{\hat{W}_l^\dagger} 
\end{align}
In order to finally compute the Kraus operators, we now need to find the matrix elements of $\hat{U}$. Since the beam splitter conserves photon number it is convenient to expand $\hat{U}$ as
\begin{align}
	\hat{U} &= \sum_{N=0}^\infty \sum_{n,m=0}^N \ket{N-m,m}\bra{N-m,m} \hat{U} \ket{N-n,n}\bra{N-n,n} \\
	&= \sum_{N=0}^\infty \sum_{n,m=0}^N U^{(N)}_{m,n} \ket{N-m}_a\ket{m}_e\bra{N-n}_a\bra{n}_e.
\end{align}
We first compute the action of $\hat{U}$ on the state with $N-n$ photons in mode $a$ and $n$ photons in mode $e$ using the Fock state definition $\ket{n}=\frac{\hat{a}^{\dagger n}}{\sqrt{n!}}\ket{0}$:
\begin{align}
	& \hat{U} \ket{N-n}_a \ket{n}_e \\
	& = \hat{U} \frac{\hat{a}^{\dagger N-n}}{\sqrt{(N-n)!}} \frac{\hat{e}^{\dagger n}}{\sqrt{n!}}  \ket{0}_a \ket{0}_e \\
	&= \frac{1}{\sqrt{(N-n)!n!}} \left( t \hat{a}^\dagger + r\hat{e}^\dagger \right)^{N-n} \left( -r \hat{a}^\dagger + t\hat{e}^\dagger \right)^n  \ket{0}_a \ket{0}_e \label{eq:compare}\\
	&= \frac{1}{\sqrt{(N-n)!n!}} \sum_{i=0}^{N-n} \sum_{j=0}^n {N-n \choose i} {n \choose j} \left( t \hat{a}^\dagger \right)^i \left(r\hat{e}^\dagger \right)^{N-n-i} \left( -r \hat{a}^\dagger\right)^j \left(t\hat{e}^\dagger \right)^{n-j}  \ket{0}_a  \ket{0}_e \\
	&= \frac{r^{N-n} t^n}{\sqrt{(N-n)!n!}} \sum_{i=0}^{N-n} \sum_{j=0}^n {N-n \choose i} {n \choose j} (-1)^j r^{j-i} t^{i-j}  \hat{a}^{\dagger i+j} \hat{e}^{\dagger N-i-j} \ket{0}_a  \ket{0}_e \\
	&= \frac{r^{N-n} t^n}{\sqrt{(N-n)!n!}} \sum_{i=0}^{N-n} \sum_{j=0}^n {N-n \choose i} {n \choose j} (-1)^j \left(\frac{r}{t}\right)^{j-i} \sqrt{(i+j)!(N-i-j)!} \ket{i+j}_a \ket{N-i-j}_e 
\end{align}
Now we can compute the matrix element 
\begin{align}
	& \bra{N-m,m} \hat{U} \ket{N-n,n}  \\
	&= \frac{r^{N-n} t^n}{\sqrt{(N-n)!n!}} \sum_{i=0}^{N-n} \sum_{j=0}^n {N-n \choose i} {n \choose j} (-1)^j \left(\frac{r}{t}\right)^{j-i} \sqrt{(i+j)!(N-i-j)!} \delta_{i+j,N-m} \delta_{m,N-i-j}  \\
%	&= \frac{r^{N-n} t^n}{\sqrt{(N-n)!n!}} \sum_{i=0}^{N-n} {N-n \choose i} {n \choose N-m-i} (-1)^{N-m-i} \left(\frac{r}{t}\right)^{N-m-2i} \sqrt{(N-m)!m!} \\
	&= \frac{r^{N-n} t^n}{\sqrt{(N-n)!n!}} \sum_{j=0}^n {N-n \choose N-m-j} {n \choose j} (-1)^j \left(\frac{r}{t}\right)^{2j-N+m)} \sqrt{(N-m)!m!} \\
\end{align}
This can be simplified to
\begin{align}
	U^{(N)}_{m,n}	&=\bra{N-m,m} \hat{U} \ket{N-n,n} \\ 
	&= \frac{\sqrt{(N-m)!m!}}{\sqrt{(N-n)!n!}} t^N \left(\frac{r}{t} \right)^{m-n} {N-n \choose N-m} \, _2F_1(-n,m-N,1+m-n,-\frac{r^2}{t^2}) \\ % \frac{n!(m-n)!}{m!} t^{-2n} P^{m-n,N-n-m}_n(t^2-r^2) = _2F_1(-n,m-N,1+m-n,\frac{-r^2}{t^2})  % 
	&= \frac{\sqrt{(N-m)!m!}}{\sqrt{(N-n)!n!}} t^N \left(\frac{r}{t} \right)^{m-n} {N-n \choose N-m} \frac{n!(m-n)!}{m!} t^{-2n} P^{(m-n,N-n-m)}_n(t^2-r^2) \\
	&= \frac{\sqrt{(N-m)!m!}}{\sqrt{(N-n)!n!}} t^N \left(\frac{r}{t} \right)^{m-n} \frac{(N-n)!}{(N-m)!(m-n)!} \frac{n!(m-n)!}{m!} t^{-2n} P^{(m-n,N-n-m)}_n(t^2-r^2) \\
	&= \frac{\sqrt{(N-n)!n!}}{\sqrt{(N-m)!m!}} t^{N-2n} \left(\frac{r}{t} \right)^{m-n} P^{(m-n,N-n-m)}_n(t^2-r^2),
\end{align}
where $P_n^{(a,b)}(x)$ are the Jacobi polynomials, which are directly related to the hypergeometric function $\, _2F_1(a,b,c,z)$. Thus, we have found the full matrix representation of $\hat{U}$ in the Fock basis
\begin{equation}
	\hat{U} = \sum_{N=0}^\infty \sum_{n,m=0}^N U^{(N)}_{m,n} \ket{N-m}_a \ket{m}_e\bra{N-n}_a \bra{n}_e,
\end{equation}
with $U^{(N)}_{m,n} =\frac{\sqrt{(N-n)!n!}}{\sqrt{(N-m)!m!}} t^{N-2n} \left(\frac{r}{t} \right)^{m-n} P^{(m-n,N-n-m)}_n(t^2-r^2)$. Now, we can read off the element
\begin{equation}
	\bra{l}_e \hat{U} \ket{0}_e = \sum_{N=0}^\infty \sum_{n,m=0}^N U^{(N)}_{m,n} \ket{N-m}_a \bra{N-n}_a  \delta_{l,m} \delta_{0,n}.
\end{equation}
Therefore, we obtain the Kraus operator
\begin{align}
	\hat{W}_{l}&= \bra{l}_e \hat{U} \ket{0}_e \\
	&= \sum_{N=0}^\infty \sum_{n,m=0}^N U^{(N)}_{m,n} \ket{N-m} \bra{N-n}  \delta_{l,m} \delta_{0,n} \\
	&= \sum_{N=l}^\infty U^{(N)}_{l,0} \ket{N-l} \bra{N} \\
	&= \sum_{N=0}^\infty U^{(N+l)}_{l,0} \ket{N} \bra{N+l} \\
	&= \sum_{N=0}^\infty \frac{\sqrt{(N+l)!}}{\sqrt{N!l!}} t^{(N+l)} \left(\frac{r}{t} \right)^{l} \underbrace{P^{(l,N)}_0(t^2-r^2)}_{=1} \ket{N} \bra{N+l} \\
	&= \left(\frac{r}{t} \right)^{l} \frac{t^l}{\sqrt{l!}} \sum_{N=0}^\infty t^{N}  \ket{N} \bra{N} \hat{a}^l\\
	&= \frac{r^l}{\sqrt{l!}} t^{\hat{n}} \hat{a}^l.
\end{align}
Thus, we have the result 
\begin{equation}
	\hat{W}_l =  \frac{r^l}{\sqrt{l!}} t^{\hat{n}} \hat{a}^l.
\end{equation}

\section{MZI Theory - coherent states}\label{app:MZI_coherent}
We start with a single-mode coherent state $\hat{\rho}=\ket{\sqrt{2}\alpha,0}\bra{\sqrt{2}\alpha,0}$ at the input of a 50:50 beamsplitter. After the beamsplitter we observe the state $\hat{\rho}=\ket{\alpha,\alpha}\bra{\alpha,\alpha}$. Then, in one arm of the MZI we apply a phase-shift and in the other we perform a memory storage and retrieval operation, $\hat{\rho}\rightarrow \ket{\hat{M} \alpha,e^{i\phi}\alpha}\bra{\hat{M} \alpha,e^{i\phi}\alpha}$, where $\ket{\hat{M}\alpha}\bra{\hat{M}\alpha}$ is the retrieved state, which we compute in the following way. For now we disregard the phase-shift arm of the MZI and add the internal memory and late time-bin state. Before the memory operation we thus have the state
\begin{equation}
	\hat{\rho} = \ket{\alpha,0,0}\bra{\alpha,0,0},
\end{equation}
where $\ket{\alpha} = e^{-|\alpha|^2/2} \sum_{n=0}^\infty \frac{\alpha^n}{\sqrt{n!}} \ket{n}$ is the coherent state in the early timebin $\hat{e}^\dagger$. The internal memory state and the output late timebin are initialized in the vacuum state $\ket{0}$. The first operation is the read-in transformation, which is modeled as a beamsplitter transformation $\hat{S}_{\mathrm{in}}$ between the early timebin $\hat{e}^\dagger$ and the internal memory mode $\hat{s}^\dagger$. Writing the coherent state as 
\begin{equation}
	\ket{\alpha} = \hat{D}(\alpha) \ket{0} = e^{\alpha \hat{e}^\dagger-\alpha^*\hat{e}} \ket{0},
\end{equation}
allows us to compute the action of the read-in beamsplitter transformation
\begin{align}
	\hat{S}_{\mathrm{in}} \ket{\alpha,0,0} &= \hat{S}_1  e^{\alpha \hat{e}^\dagger-\alpha^*\hat{e}} \ket{0,0,0} \\
	&= \hat{S}_{\mathrm{in}} e^{\alpha \hat{e}^\dagger-\alpha^*\hat{e}}\hat{S}_{\mathrm{in}}^\dagger \underbrace{\hat{S}_{\mathrm{in}} \ket{0,0,0}}_{=\ket{0,0,0}} \\
	&= e^{\alpha (\sqrt{t_{\mathrm{in}}}\hat{e}^\dagger+\sqrt{r_{\mathrm{in}}}\hat{s}^\dagger)-\alpha^*(\sqrt{t_{\mathrm{in}}}\hat{e}+\sqrt{r_{\mathrm{in}}}\hat{s})} \ket{0,0,0} \\
	&= e^{\sqrt{t_{\mathrm{in}}}(\alpha \hat{e}^\dagger - \alpha^*\hat{e})+ \sqrt{r_{\mathrm{in}}} (\alpha\hat{s}^\dagger-\alpha^*\hat{s})} \ket{0,0,0} \\
	&= e^{\sqrt{t_{\mathrm{in}}}(\alpha \hat{e}^\dagger - \alpha^*\hat{e})} e^{\sqrt{r_{\mathrm{in}}} (\alpha\hat{s}^\dagger-\alpha^*\hat{s})} \ket{0,0,0} \\
	&= \ket{\sqrt{t_{\mathrm{in}}}\alpha,\sqrt{r_{\mathrm{in}}}\alpha,0}. \\
\end{align}
The read-out transformation $\hat{S}_{\mathrm{out}}$ between the internal state $\hat{s}^\dagger$ and the late timebin $\hat{l}^\dagger$ is applied in the same way
\begin{align}
	\hat{S}_2 \ket{\sqrt{t_{\mathrm{in}}}\alpha,\sqrt{r_{\mathrm{in}}}\alpha,0} &= \ket{\sqrt{t_{\mathrm{in}}}\alpha,\sqrt{r_{\mathrm{in}}t_{\mathrm{out}}}\alpha,\sqrt{r_{\mathrm{in}}}\sqrt{r_{\mathrm{out}}}\alpha}, \\
\end{align}
as we cannot distinguish between the read-in and read out efficiency in experiment, we assume $t_{\mathrm{in}} = t_{\mathrm{out}} = t$ and $r_{\mathrm{in}} = r_{\mathrm{out}} = r$ thus, where $t = 1- \sqrt{\eta_{\mathrm{int}}}$ and $r = \sqrt{\eta_{\mathrm{int}}}$. Substituting for $r$ and $t$ leaves us with the intermediate state
\begin{equation}
	\hat{\rho}=\ket{\sqrt{t}\alpha,\sqrt{rt}\alpha,r\alpha}\bra{\sqrt{t}\alpha,\sqrt{rt}\alpha,r\alpha}.
\end{equation}
We now trace out the internal memory state and the early timebin and obtain the single-mode coherent state
\begin{equation}
	\hat{\rho}=\ket{r\alpha}\bra{r\alpha}.
\end{equation}
As the next step we apply the Krauss operators of the lossy beamsplitter channel
\begin{align}
	\hat{A}_k \ket{r\alpha} &= p_k \tau^{\hat{n}/2}\hat{l}^k \ket{r\alpha} \\
	&= p_k \tau^{\hat{n}/2}(r\alpha)^k \ket{r \alpha}\\
	&= e^{-r^2|\alpha|^2/2} p_l (r\alpha)^k \tau^{\hat{n}/2} \sum_{n=0}^\infty \frac{(r \alpha)^n}{\sqrt{n!}} \ket{n}\\
	&= e^{-r^2|\alpha|^2/2} p_k (r\alpha)^k  \sum_{n=0}^\infty \frac{(r \alpha)^n}{\sqrt{n!}} \tau^{n/2}\ket{n}\\
	&= e^{-r^2|\alpha|^2/2} p_k (r\alpha)^k  \sum_{n=0}^\infty \frac{(r \sqrt{\tau} \alpha)^n}{\sqrt{n!}} \ket{n}\\
	&= e^{-r^2|\alpha|^2/2} p_k (r\alpha)^k  e^{\tau r^2 |\alpha|^2/2} \ket{\sqrt{\tau}r\alpha } \\
	&= p_k (r\alpha)^k  e^{(\tau-1)r^2|\alpha|^2/2} \ket{\sqrt{\tau}r\alpha }.
\end{align}
Thus the density matrix is transformed to
\begin{align}
	\sum_{k=0}^\infty \hat{A}_k \ket{r\alpha}\bra{r\alpha} \hat{A}_k^\dagger &= \underbrace{\sum_{k=0}^\infty  p_k^2 (r^2|\alpha|^2)^k}_{\mathclap{=\sum_{k=0}^\infty \frac{((1-\tau) r^2|\alpha|^2)^k}{k!} = e^{(1-\tau)r^2|\alpha|^2}}}  e^{(\tau-1)r^2|\alpha|^2} \ket{\sqrt{\tau}r\alpha }\bra{\sqrt{\tau}r\alpha } \\
	&= \ket{\sqrt{\tau}r\alpha }\bra{\sqrt{\tau}r\alpha },
\end{align}
where $\tau = \kappa/G$ and $G=1+(1-\kappa)\bar{n}_\mathrm{B}$. So far, the memory read-in, read-out and loss-channel have only attenuated the input coherent state from $\ket{\alpha}$ to $\ket{\sqrt{\tau}r\alpha}$ and we define for the sake of brevity $\beta = \sqrt{\tau}r\alpha$. To simulate quantum memories we set the beamsplitter transmissivity $\kappa = \eta_{\mathrm{trans}}$ and $\bar{n}_{\mathrm{B}} = \braket{n}_{\mathrm{noise}}/(1-\eta_{\mathrm{trans}})$ and apply the action of a quantum-limited amplifier channel:

\begin{align}
	\hat{B}_k \ket{\beta} &= q_k \hat{l}^{\dagger k} G^{-\hat{n}/2}\ket{\beta} \\
	&= q_k \hat{l}^{\dagger k} e^{-|\beta|^2/2} G^{-\hat{n}/2} \sum_{n=0}^\infty \frac{\beta^n}{\sqrt{n!}} \ket{n} \\
	&= q_k \hat{l}^{\dagger k} e^{-|\beta|^2/2} \sum_{n=0}^\infty \frac{(G^{-1/2}\beta)^n}{\sqrt{n!}} \ket{n} \\
	&= q_k  e^{-|\beta|^2/2} \sum_{n=0}^\infty \frac{(G^{-1/2}\beta)^n}{\sqrt{n!}} \hat{l}^{\dagger k} \ket{n} \\
	&= q_k  e^{-|\beta|^2/2} \sum_{n=0}^\infty \frac{(G^{-1/2}\beta)^n}{\sqrt{n!}} \frac{\sqrt{(n+k)!}}{\sqrt{n!}} \ket{n+k}.
\end{align}
The state thus transforms to
\begin{align}
	\hat{\rho} & \rightarrow \sum_{k=0}^\infty \hat{B}_k \ket{\beta}\bra{\beta} \hat{B}_k^\dagger \\
	&= e^{-|\beta|^2}\sum_{k,n,m=0}^\infty q_k^2 \frac{(G^{-1/2}\beta)^n(G^{-1/2}\beta^*)^m}{n!m!} \sqrt{(n+k)!(m+k)!} \ket{n+k}\bra{m+k}.
\end{align}
The matrix elements in the Fock-basis are thus
\begin{align}
	\rho_{p,q} & = \bra{p}\hat{\rho}\ket{q} \\
	&= e^{-|\beta|^2}\sum_{k=0}^\infty q_k^2 \frac{(G^{-1/2}\beta)^{p-k} (G^{-1/2}\beta^*)^{q-k} \sqrt{p!q!}}{(p-k)! (q-k)!} \\
	&= e^{-|\beta|^2}\sum_{k=0}^{\min(p,q)} q_k^2 \frac{(G^{-1/2}\beta)^{p-k} (G^{-1/2}\beta^*)^{q-k} \sqrt{p!q!}}{(p-k)! (q-k)!} \\
	&= G^{-1} e^{-|\beta|^2}\sum_{k=0}^{\min(p,q)} \left(\frac{G-1}{G}\right)^k \frac{(G^{-1/2}\beta)^{p-k} (G^{-1/2}\beta^*)^{q-k} \sqrt{p!q!}}{k!(p-k)! (q-k)!}.
	\label{eq:memory_output}
\end{align}
To resolve the $\min(p,q)$ expression we take a look at the elements $\rho_{n,n+m}$ ($n,m\ge 0$)
\begin{align}
	\rho_{n,n+m} & = \frac{e^{-|\beta|^2}}{G} \sum_{k=0}^{n} \left(\frac{G-1}{G}\right)^k \frac{(G^{-1/2}\beta)^{n-k} (G^{-1/2}\beta^*)^{n+m-k}  \sqrt{n!(n+m)!}}{k!(n-k)! (n+m-k)!}.
\end{align}
Using Mathematica we rewrite this as
\begin{align}
	\rho_{n,n+m} & = \frac{e^{-|\beta|^2}}{G} \left(\frac{G-1}{G}\right)^n (-1)^n \frac{\left( G^{-1/2} \beta^*\right)^m}{\sqrt{n!(n+m)!}} U\left(-n,m+1,\frac{|\beta|^2}{1-G}\right) ,
\end{align}
where $U(a,b,z)$ is the confluent hypergeometric function. For integer $a,b$ this can be expressed in terms of the generalized Laguerre polynomials using the relation $L_n^{(\alpha)}(x)=\frac{(-1)^n}{n!}U(-n,\alpha+1,x)$
\begin{equation}
	\rho_{n,n+m} = \frac{e^{-|\beta|^2}}{G} \left(\frac{G-1}{G}\right)^n \frac{\sqrt{n!}\left( G^{-1/2} \beta^*\right)^m}{\sqrt{(n+m)!}} L_n^{(m)}\left(\frac{|\beta|^2}{1-G}\right),
\end{equation}
with $L_n^{(m)}(x) = \sum_{k=0}^n (-1)^k \binom{n+m}{n-k} \frac{x^k}{k!}$. The remaining elements of $\hat{\rho}$ are obtained via complex conjugation $\rho_{n+m,n} = \rho_{n,n+m}^*$. In summary, the retrieved state from the quantum memory reads as
\begin{equation}
	\ket{\hat{M}\alpha}\bra{\hat{M}\alpha} = \sum_{p,q=0}^\infty \rho_{p,q} \ket{p}\bra{q},
\end{equation}
with $\rho_{p,q}$ given in Eq.~(\ref{eq:memory_output}). \par
Going back to the MZI state we now have the complete state
\begin{equation}
	\hat{\rho} = \sum_{p,q=0}^{\infty} \rho_{p,q} \ket{p,e^{i\phi}\alpha}\bra{q,e^{i\phi}\alpha},
\end{equation}
after the memory and phase shift operation. This state is now recombined on a second 50:50 beamsplitter $\hat{\rho} \rightarrow \hat{U}\hat{\rho}\hat{U}^\dagger$. However, going forward as before produces quite large and tedious expressions and thus we will slightly switch gears and perform an equivalent calculation. To proceed we first introduce the observable we seek to measure, namely the photon number operator at one of the outputs of the MZI, say $\hat{n}_a = \hat{a}^\dagger \hat{a}$. Formally, we obtain the expectation value of $\hat{n}_a$ from 
\begin{equation}
	\left\langle \hat{n}_a \right\rangle_{\hat{U},\hat{\rho}} = \Tr(\hat{a}^\dagger\hat{a}\hat{U}\hat{\rho}\hat{U}^\dagger),
\end{equation}
where $\hat{U}$ is the 50:50 beamsplitter transformation. Using the cyclical property of the trace, we also find 
\begin{equation}
	\left\langle \hat{n}_a \right\rangle_{\hat{U},\hat{\rho}}  = \Tr(\hat{U}^\dagger \hat{a}^\dagger \hat{a}\hat{U}\hat{\rho}).
\end{equation}
Thus, instead of applying the beamsplitter transformation directly onto the (complicated) state $\hat{\rho}$, we apply the inverse 50:50 beamsplitter transformation on the observable and will obtain the same result. We obtain firstly
\begin{align}
	\hat{U}^\dagger \hat{a}^\dagger \hat{a} \hat{U} &= \hat{U}^\dagger \hat{a}^\dagger \hat{U} \hat{U}^\dagger \hat{a} \hat{U}\\
	&=  \frac{1}{2} \left(\hat{a}^\dagger-\hat{b}^\dagger \right) \left(\hat{a}-\hat{b} \right)\\ 
	&= \frac{1}{2}\left( \hat{a}^\dagger \hat{a} + \hat{b}^\dagger\hat{b}-\left(\hat{a}^\dagger\hat{b}+\hat{a}\hat{b}^\dagger \right) \right)
\end{align}
and thus
\begin{align}
	\left\langle \hat{n}_a \right\rangle_{\hat{U},\hat{\rho}} &= \frac{1}{2}\Tr(\left( \hat{a}^\dagger \hat{a} + \hat{b}^\dagger\hat{b}-\left(\hat{a}^\dagger\hat{b}+\hat{a}\hat{b}^\dagger \right) \right) \hat{\rho})\\
	&=  \frac{1}{2}\left( \left\langle \hat{n}_a \right\rangle_{\hat{\rho}} +\left\langle \hat{n}_b \right\rangle_{\hat{\rho}} - \left\langle \hat{a}^\dagger\hat{b}+\hat{a}\hat{b}^\dagger \right\rangle_{\hat{\rho}} \right).
\end{align}
Thus we need to compute the average photon number in path $a$ and $b$ and the expectation value of the coupling term $\hat{a}^\dagger\hat{b}+\hat{a}\hat{b}^\dagger$ before the final 50:50 beam splitter. We proceed accordingly
\begin{align}
	\left\langle \hat{n}_a \right\rangle_{\hat{\rho}} &= \Tr(\hat{n}_a\hat{\rho}) \\
	&= \sum_{n,m=0}^\infty \bra{n,m} \hat{a}^\dagger \hat{a} \hat{\rho} \ket{n,m} \\
	&= \sum_{n=0}^\infty n \bra{n}\hat{\rho}\ket{n}_a \\
	&= \sum_{n=0}^\infty n \rho_{n,n} \\
	&= \sum_{n=0}^\infty n \frac{e^{-|\beta|^2}}{G} \left(\frac{G-1}{G}\right)^n L_n^{(0)}\left(\frac{|\beta|^2}{1-G}\right) \\
	&= \frac{e^{-|\beta|^2}}{G} \sum_{n=0}^\infty \left(\frac{G-1}{G}\right)^n n L_n^{(0)}\left(\frac{|\beta|^2}{1-G}\right) \eqqcolon \gamma.
	\label{eq:gamma}
\end{align}
This expression is not further Mathematica simplify-able but it converges for any given $\beta$ and $G>1$ and can be computed efficiently. The next term is simply 
\begin{equation}
	\left\langle \hat{n}_b \right\rangle_{\hat{\rho}} = |\alpha|^2.
\end{equation}
Finally, we need to compute the expression
\begin{align}
	\left\langle \hat{a}^\dagger\hat{b}+\hat{a}\hat{b}^\dagger \right\rangle_{\hat{\rho}} &= \sum_{n,m=0}^\infty \sum_{p,q=0}^\infty \rho_{p,q} \bra{n,m} \left(\hat{a}^\dagger\hat{b}+\hat{a}\hat{b}^\dagger \right) \ket{p,e^{i\phi}\alpha}\left\langle q,e^{i\phi}\alpha|n,m \right\rangle \\
	&= e^{-|\alpha|^2/2}\sum_{n,m=0}^\infty \sum_{p,q=0}^\infty \rho_{p,q} \bra{n,m} \left(\hat{a}^\dagger\hat{b}+\hat{a}\hat{b}^\dagger \right) \ket{p,e^{i\phi}\alpha} \delta_{q,n} \frac{(e^{-i\phi}\alpha^*)^m}{\sqrt{m!}} \\
	\begin{split}
		&= e^{-|\alpha|^2/2} \sum_{n,m=0}^\infty \sum_{p,q=0}^\infty \rho_{p,q} \left[\sqrt{n(m+1)} \bra{n-1,m+1}+\sqrt{(n+1)m}\bra{n+1,m-1} \right] \\
		& \qquad \ket{p,e^{i\phi}\alpha} \delta_{q,n} \frac{(e^{-i\phi}\alpha^*)^m}{\sqrt{m!}} 
	\end{split}	\\
	\begin{split}
		&= e^{-|\alpha|^2}\sum_{n,m=0}^\infty \sum_{p,q=0}^\infty \rho_{p,q} \left[\sqrt{n(m+1)}\delta_{p,n-1} \frac{(e^{i\phi} \alpha)^{m+1}}{\sqrt{(m+1)!}}+\sqrt{(n+1)m}\delta_{p,n+1} \frac{(e^{i\phi} \alpha)^{m-1}}{\sqrt{(m-1)!}} \right] \\
		& \qquad \delta_{q,n} \frac{(e^{-i\phi}\alpha^*)^m}{\sqrt{m!}} 
	\end{split}	\\
	\begin{split}
		&= e^{-|\alpha|^2} \sum_{n,m=0}^\infty \sum_{p,q=0}^\infty \frac{(e^{i\phi} \alpha)^{m+1}}{\sqrt{(m+1)!}}\frac{(e^{-i\phi}\alpha^*)^m}{\sqrt{m!}} \sqrt{n(m+1)} \rho_{p,q}  \delta_{p,n-1}\delta_{q,n}  \\
		& + \sum_{n,m=0}^\infty \sum_{p,q=0}^\infty \frac{(e^{i\phi} \alpha)^{m-1}}{\sqrt{(m-1)!}}\frac{(e^{-i\phi}\alpha^*)^m}{\sqrt{m!}} \sqrt{n(m+1)} \rho_{p,q}  \delta_{p,n+1}\delta_{q,n}  
	\end{split}	\\
	&= e^{-|\alpha|^2} \sum_{n,m=0}^\infty \sqrt{n} \rho_{n-1,n} \frac{|\alpha|^{2m}}{m!}  \alpha e^{i\phi} + \sum_{n,m=0}^\infty \sqrt{n+1} \rho_{n+1,n} \frac{|\alpha|^{2(m-1)}}{(m-1)!} \alpha^* e^{-i\phi} \\
	&= \left( \alpha e^{i\phi} \sum_{n=0}^\infty \sqrt{n} \rho_{n-1,n}  + \alpha^* e^{-i\phi} \sum_{n=0}^\infty \sqrt{n+1} \rho_{n+1,n}  \right)\\
	&= \left( \alpha e^{i\phi} \underbrace{\sum_{n=0}^\infty \sqrt{n+1} \rho_{n,n+1}}_{\eqcolon \xi }  + \alpha^* e^{-i\phi} \sum_{n=0}^\infty \sqrt{n+1} \rho^*_{n,n+1}  \right)\\
	&= \left( \alpha e^{i\phi} \xi + \alpha^* e^{-i\phi} \xi^*\right) ,
\end{align}
with 
\begin{align}
	\xi & = \sum_{n=0}^\infty \sqrt{n+1} \rho_{n,n+1}\\
	&= \sum_{n=0}^\infty \sqrt{n+1} \frac{e^{-|\beta|^2}}{G} \left(\frac{G-1}{G}\right)^n \frac{\sqrt{n!}\left( G^{-1/2} \beta^*\right)}{\sqrt{(n+1)!}} L_n^{(1)}\left(\frac{|\beta|^2}{1-G}\right)\\
	&= \frac{e^{-|\beta|^2}}{G} \left( G^{-1/2} \beta^*\right) \sum_{n=0}^\infty \left(\frac{G-1}{G}\right)^n L_n^{(1)}\left(\frac{|\beta|^2}{1-G}\right),
	\label{eq:xi}
\end{align}
which also converges and can be computed for any given $\beta$ and $G>1$. Collecting all these results we have
\begin{equation}
	\left\langle \hat{n}_a \right\rangle_{\hat{U},\hat{\rho}} = \frac{1}{2} \left(\gamma + |\alpha|^2 - \left( \alpha e^{i\phi} \xi + \alpha^* e^{-i\phi} \xi^*\right) \right) = \frac{1}{2}\left(\gamma +|\alpha|^2  -2\alpha\xi \cos(\phi) \right),
\end{equation}
where we assumed $\alpha$ and $\xi$ to be real and positive. The fringe visibility is thus 
\begin{align}
	\nu &= \frac{2\alpha\xi}{\gamma + |\alpha|^2},
\end{align}
where $\sqrt{2}\alpha$ is the coherent amplitude of the input state of the MZI, $\xi=\xi(\beta,G)$ (see Eq.~(\ref{eq:xi})) and $\gamma=\gamma(\beta,G)$ (see Eq.~(\ref{eq:gamma})) are analytic functions of the memory-attenuated amplitude $\beta = \sqrt{\tau} r \alpha$ and the gain parameter $G$.
\section{Single photon MZI}
We start with the single-mode single photon state
\begin{equation}
	\hat{\rho} = \ket{1,0}\bra{1,0},
\end{equation}
which gets split on the first 50:50 beam splitter
\begin{equation}
	\hat{\rho} \rightarrow \frac{1}{2}\left(\ket{1,0}+\ket{0,1} \right)\left(\bra{1,0}+\bra{0,1} \right).
\end{equation}
Since this is still a pure state, we use the shorthand notation
\begin{equation}
	\hat{\rho} = \frac{1}{\sqrt{2}}\left(\ket{1,0}+\ket{0,1} \right)\left(h.c.\right).
\end{equation}
The phaseshift transformation yields
\begin{equation}
	\hat{\rho} = \frac{1}{\sqrt{2}}\left(\ket{1,0}+e^{i\phi}\ket{0,1} \right)\left(h.c.\right).
\end{equation}
We now add the internal state and late time-bin
\begin{equation}
	\hat{\rho} = \frac{1}{\sqrt{2}}\left(\ket{1,0,0,0}+e^{i\phi}\ket{0,0,0,1} \right)\left(h.c.\right),
\end{equation}
and apply the storage operation
\begin{equation}
	\hat{\rho} = \frac{1}{\sqrt{2}}\left(\sqrt{t}\ket{1,0,0,0}+\sqrt{r}\ket{0,1,0,0}+e^{i\phi}\ket{0,0,0,1} \right)\left(h.c.\right),
\end{equation}
and read-out operation
\begin{equation}
	\hat{\rho} = \frac{1}{\sqrt{2}}\left(\sqrt{t}\ket{1,0,0,0}+\sqrt{rt}\ket{0,1,0,0}+r\ket{0,0,1,0}+e^{i\phi}\ket{0,0,0,1} \right)\left(h.c.\right).
\end{equation}
When we now trace out the early time-bin we will obtain a mixed state
\begin{equation}
	\Tr_1(\hat{\rho}) = \frac{t}{2}\ket{0,0,0}\bra{0,0,0}+\frac{1}{\sqrt{2}}\left(\sqrt{rt}\ket{1,0,0}+r\ket{0,1,0}+e^{i\phi}\ket{0,0,1} \right)\left(h.c.\right).
\end{equation}
We further trace out the internal state
\begin{equation}
	\Tr_2(\Tr_1(\hat{\rho})) = \frac{t}{2}\ket{0,0}\bra{0,0}+\frac{rt}{2}\ket{0,0}\bra{0,0}+\frac{1}{\sqrt{2}}\left(r\ket{1,0}+e^{i\phi}\ket{0,1} \right)\left(h.c.\right).
\end{equation}
The remaining complete state is thus
\begin{equation}
	\hat{\rho} = \frac{1-r^2}{2}\ket{0,0}\bra{0,0}+\frac{1}{\sqrt{2}}\left(r\ket{1,0}+e^{i\phi}\ket{0,1} \right)\left(h.c.\right).
\end{equation}
We apply the loss operators
\begin{align}
	\hat{\rho} &\rightarrow \sum_{l=0}^\infty \hat{A}_l \hat{\rho} \hat{A}_l^\dagger \\
	&= \underbrace{p_0^2}_{=1}\tau^{\frac{\hat{n}}{2}} \left(\frac{1-r^2}{2}\ket{0,0}\bra{0,0}+\frac{1}{\sqrt{2}}\left(r\ket{1,0}+e^{i\phi}\ket{0,1} \right)\left(h.c.\right) \right) \tau^{\frac{\hat{n}}{2}} + \underbrace{p_1^2}_{=1-\tau} \tau^{\frac{\hat{n}}{2}} \frac{r^2}{2} \ket{0,0} \bra{0,0}\tau^{\frac{\hat{n}}{2}} \\
	&= \frac{1-r^2}{2}  \ket{0,0}\bra{0,0}+ \frac{1}{\sqrt{2}}\left(r\sqrt{\tau}\ket{1,0}+e^{i\phi}\ket{0,1} \right)\left(h.c.\right) + \frac{(1-\tau)r^2}{2} \ket{0,0} \bra{0,0} \\
	&= \frac{1-\tau r^2}{2} \ket{0,0}\bra{0,0} + \frac{1}{\sqrt{2}}\left(r\sqrt{\tau}\ket{1,0}+e^{i\phi}\ket{0,1} \right)\left(h.c.\right).
\end{align}
Now we apply the noise operators step by step. We first analyze the vacuum term
\begin{equation}
	\hat{B}_k \ket{0} = q_k \left(\hat{a}^\dagger\right)^k G^{-\hat{n}/2} \ket{0} = q_k \sqrt{k!} \ket{k},
\end{equation}
and therefore
\begin{equation}
	\sum_{k=0}^\infty \hat{B}_k \ket{0,0} \bra{0,0} \hat{B}_k^\dagger = \sum_{k=0}^\infty q_k^2 k! \ket{k,0}\bra{k,0}.
\end{equation}
For the remaining term we also need
\begin{equation}
	\hat{B}_k \ket{1} = q_k \left(\hat{a}^\dagger\right)^k G^{-\hat{n}/2} \ket{1} = \frac{1}{\sqrt{G}} q_k \sqrt{(k+1)!} \ket{k+1},
\end{equation}
and therefore
\begin{equation}
	\hat{B}_k \frac{1}{\sqrt{2}}\left(r\sqrt{\tau}\ket{1,0}+e^{i\phi}\ket{0,1} \right)\left(h.c.\right) = \frac{1}{\sqrt{2}}\left(r\sqrt{\frac{\tau}{G}} q_k \sqrt{(k+1)!}  \ket{k+1,0}+e^{i\phi} q_k \sqrt{k!} \ket{k,1} \right)\left(h.c.\right).
\end{equation}
In summary, after the complete memory operation we have the state
\begin{align}
	\begin{split}
		\hat{\rho} = & \sum_{k=0}^\infty  \frac{1-\tau r^2}{2} q_k^2 k! \ket{k,0}\bra{k,0} + \frac{1}{2} \Big[ r^2\frac{\tau}{G} q_k^2 (k+1)!\ket{k+1,0}\bra{k+1,0} + q_k^2 k! \ket{k,1}\bra{k,1} \\
		& + r\sqrt{\frac{\tau}{G}} q_k^2 \sqrt{(k+1)!k!}e^{-i\phi}\ket{k+1,0}\bra{k,1} + r\sqrt{\frac{\tau}{G}} q_k^2 \sqrt{(k+1)!k!}e^{i\phi} \ket{k,1}\bra{k+1,0} \Big],
	\end{split}	
\end{align}
with the elements
\begin{align}
	\rho_{\substack{n,m\\ p,q}} &= \bra{n,m} \hat{\rho} \ket{p,q} \\
	\begin{split}
		&= \frac{1-\tau r^2}{2} q_n^2 n! \delta_{n,p} \delta_{m,0} \delta_{q,0} + \frac{1}{2} \Big[ r^2\frac{\tau}{G} q_{n-1}^2 n!\delta_{n,p} \delta_{m,0} \delta_{q,0}  + q_n^2 n! \delta_{n,p} \delta_{m,1} \delta_{q,1}  \\
		& + r\sqrt{\frac{\tau}{G}} q_{n-1}^2 \sqrt{n!(n-1)!}e^{-i\phi}\delta_{p,n-1} \delta_{m,0} \delta_{q,1}  + r\sqrt{\frac{\tau}{G}} q_n^2 \sqrt{(n+1)!n!}e^{i\phi} \delta_{p,n+1} \delta_{m,1} \delta_{q,0}  \Big].
	\end{split}	
\end{align}
Now we need to compute the expectation value 
\begin{align}
	\left\langle \hat{n}_a \right\rangle_{\hat{\rho}} &= \sum_{n,m=0}^\infty n \rho_{\substack{n,m\\ n,m}} \\
	&= \sum_{n=0}^\infty n \left[\frac{(1-\tau r^2)}{2}q_n^2n! +\frac{r^2 \tau }{2G} q_{n-1}^2 n! + \frac{1}{2} q_n^2 n! \right]\\
	&= \left(1-\frac{\tau r^2}{2}\right) \sum_{n=0}^\infty n q_n^2n! + \frac{r^2 \tau }{2G}\sum_{n=0}^\infty n q_{n-1}^2 n! \\
	&= \left(1-\frac{\tau r^2}{2}\right) \frac{1}{G} \underbrace{\sum_{n=0}^\infty n \left(\frac{G-1}{G}\right)^n}_{=G(G-1)} + \frac{r^2 \tau }{2G(G-1)}\underbrace{\sum_{n=0}^\infty n^2  \left(\frac{G-1}{G}\right)^n}_{=G(G-1)(2G-1)}\\	
	&= \left(1-\frac{\tau r^2}{2}\right) (G-1) + \frac{r^2 \tau }{2}(2G-1)\\
	&= G\left(1+\frac{\tau r^2}{2}\right) -1.
\end{align}
Similarly we obtain
\begin{align}
	\left\langle \hat{n}_b \right\rangle_{\hat{\rho}} &= \sum_{n,m=0}^\infty m \rho_{\substack{n,m\\ n,m}} \\
	&= \sum_{n,m=0}^\infty m \frac{1}{2} q_n^2 n! \delta_{n,n}\delta_{m,1} \delta_{m,1}\\
	&= \frac{1}{2} \sum_{n=0}^\infty q_n^2 n! \\
	&= \frac{1}{2G} \sum_{n=0}^\infty \left(\frac{G-1}{G}\right)^n  \\
	&= \frac{1}{2G} G \\
	&= \frac{1}{2}.
\end{align}
The last expectation value we need is
\begin{align}
	\left\langle \hat{a}^\dagger \hat{b} + \hat{a}\hat{b}^\dagger \right\rangle_{\hat{\rho}} &= \sum_{n,m=0}^\infty \bra{n,m} \left( \hat{a}^\dagger \hat{b} + \hat{a}\hat{b}^\dagger\right) \hat{\rho} \ket{n,m} \\
	&= \sum_{n,m=0}^\infty \sqrt{n(m+1)} \rho_{\substack{n-1,m+1\\ n,m}} +\sqrt{(n+1)m} \rho_{\substack{n+1,m-1\\ n,m}}\\
	&= \sum_{n,m=0}^\infty \sqrt{n(m+1)} \left[\frac{r}{2}\sqrt{\frac{\tau}{G}} q^2_{n-1}\sqrt{(n-1)!n!} e^{i\phi}\delta_{n,n}\delta_{m,0} \right]\\
	& +\sqrt{(n+1)m} \left[\frac{r}{2}\sqrt{\frac{\tau}{G}} q^2_{n}\sqrt{(n+1)!n!} e^{-i\phi}\delta_{n,n}\delta_{m,1} \right]\\
	&= \sum_{n=0}^\infty \sqrt{n} \frac{r}{2}\sqrt{\frac{\tau}{G}} q^2_{n-1}\sqrt{(n-1)!n!} e^{i\phi} + \sum_{n=0}^\infty \sqrt{n+1} \frac{r}{2}\sqrt{\frac{\tau}{G}} q^2_{n}\sqrt{(n+1)!n!} e^{-i\phi}\\
	&= \frac{r}{2}\sqrt{\frac{\tau}{G}} \left[e^{i\phi} \underbrace{\sum_{n=0}^\infty n! q^2_{n-1}}_{=\sum_{n=0}^\infty (n+1)! q^2_{n}} +  e^{-i\phi}\sum_{n=0}^\infty (n+1)!q^2_{n} \right] \\
	&= \frac{r}{2}\sqrt{\frac{\tau}{G}} 2\cos(\phi) \sum_{n=0}^\infty (n+1)!q^2_{n} \\
	&= r\sqrt{\frac{\tau}{G}} \cos(\phi) \frac{1}{G} \sum_{n=0}^\infty (n+1) \left(\frac{G-1}{G}\right)^n\\
	&= r\sqrt{\frac{\tau}{G}} \cos(\phi) \frac{1}{G} G^2\\
	&= r\sqrt{\tau G} \cos(\phi).
\end{align}
Thus we have the interference fringe at the output
\begin{equation}
	\left\langle \hat{n}_a \right\rangle_{\hat{U}\hat{\rho}\hat{U}^\dagger} = \frac{G}{2}\left(1+\frac{\tau r^2}{2}\right) -\frac{1}{4}-\frac{r\sqrt{\tau G}}{2} \cos(\phi),
\end{equation}
with the visibility
\begin{equation}
	\nu = \frac{r\sqrt{\tau G}}{G(1+\frac{\tau r^2}{2})-\frac{1}{2}}.
\end{equation}

\end{document}